%% file: paper.tex
\documentclass[sigconf]{acmart}

\copyrightyear{2022}
\acmYear{2022}
\setcopyright{acmlicensed}
\acmConference[WiSec '22] {Proceedings of the 15th ACM Conference on Security and Privacy in Wireless and Mobile Networks}{May 16--19, 2022}{San Antonio, TX, USA}
\acmBooktitle{Proceedings of the 15th ACM Conference on Security and Privacy in Wireless and Mobile Networks (WiSec '22), May 16--19, 2022, San Antonio, TX, USA}
\acmPrice{15.00}
\acmISBN{978-1-4503-9216-7/22/05}
\acmDOI{10.1145/3507657.3528539}

\usepackage{algorithmic}
\usepackage{graphicx}
\usepackage{textcomp}
\usepackage{bmpsize}
\usepackage{xcolor}
\usepackage{lipsum}

\input{include}

\newcommand{\name}{SP-MAC\xspace}
\newcommand{\dname}{R2-D2\xspace}

\usepackage{tabularx}
\usepackage{subcaption}
\usepackage[nolist,nohyperlinks]{acronym}
\usepackage{stackengine}

\newcolumntype{Y}{>{\centering\arraybackslash}X}

\usepackage[inline]{enumitem}

\begin{acronym}[TDMA]
	\acro{ICS}{Industrial Control System}
	\acro{MitM}{Man-in-the-Middle}
	\acro{WSN}{Wireless Sensor Network}
	\acro{PER}{Packet Error Rate}
	\acro{CPS}{Cyber-physical System}
	\acro{DoS}{Denial of Service}
	
	\acrodefplural{ICS}[ICSs]{Industrial Control Systems}
	\acrodefplural{WSN}[WSNs]{Wireless Sensor Networks}
	\acrodefplural{PER}[PERs]{Packet Error Rates}
	\acrodefplural{CPS}[CPSs]{Cyber-physical Systems}
\end{acronym}

\newif\ifmanualheader

\manualheaderfalse

\ifmanualheader
  \makeatletter
  \if@ACM@anonymous
  \else
    \g@addto@macro\@subtitlenotes{\global\let\authors=\cleanauthors}
  \fi
  \makeatother
\fi

\begin{document}
\fancyhead{}

\title[{Take a Bite of the Reality Sandwich: Revisiting the Security of Progressive Message Authentication Codes}]{Take a Bite of the Reality Sandwich: Revisiting the \\ Security of Progressive Message Authentication Codes}

\ifmanualheader
  \author{Eric Wagner$^{*,\dagger}$, Jan Bauer$^{*}$, and Martin Henze$^{\ddagger,*}$}
  \def\cleanauthors{Eric Wagner, Jan Bauer, and Martin Henze}
  \affiliation{
    $^*$\textit{Cyber Analysis \& Defense}, Fraunhofer FKIE \country{Germany} $\cdot$ \{firstname.lastname\}@fkie.fraunhofer.de\\
    $^\dagger$\textit{Communication and Distributed Systems}, RWTH Aachen University \country{Germany} $\cdot$ wagner@comsys.rwth-aachen.de\\ %
    $^\ddagger$\textit{Security and Privacy in Industrial Cooperation}, RWTH Aachen University \country{Germany} $\cdot$ henze@cs.rwth-aachen.de
  }
\else
  \author{Eric Wagner}
  \email{eric.wagner@fkie.fraunhofer.de}
  \affiliation{%
    \institution{Fraunhofer FKIE}
    \country{}
  }
  \affiliation{%
    \institution{RWTH Aachen University}
    \country{}
  }

  \author{Jan Bauer}
  \email{jan.bauer@fkie.fraunhofer.de}
  \affiliation{%
    \institution{Fraunhofer FKIE}
    \country{}
  }

  \author{Martin Henze}
  \email{henze@cs.rwth-aachen.de}
  \affiliation{%
	\institution{RWTH Aachen University}
    \country{}
  }
  \affiliation{%
    \institution{Fraunhofer FKIE}
    \country{}
  }
\fi

\renewcommand{\shortauthors}{\cleanauthors}



\begin{abstract}
	
	Message authentication guarantees the integrity of messages exchanged over untrusted channels.
	However, to achieve this goal, message authentication considerably expands packet sizes, which is especially problematic in constrained wireless environments.
	To address this issue, \emph{progressive} message authentication provides initially reduced integrity protection that is often sufficient to process messages upon reception.
	This reduced security is then successively improved with subsequent messages to uphold the strong guarantees of traditional integrity protection. 
	However, contrary to previous claims, we show in this paper that existing progressive message authentication schemes are highly susceptible to packet loss induced by 
	poor channel conditions or jamming attacks.
	Thus, we consider it imperative to rethink how authentication tags depend on the successful reception of surrounding packets.
	To this end, we propose \dname, which uses randomized dependencies with parameterized security guarantees to increase the resilience of progressive authentication 
	against packet loss. 
	To deploy our approach to resource-constrained devices, we introduce \name{}, which implements \dname using efficient XOR operations.
	Our evaluation shows that \name{} is resilient to sophisticated network-level attacks and operates as resources-conscious and fast as existing, yet insecure, progressive message authentication schemes.

\end{abstract}

\maketitle

\section{Introduction}
\label{sec:intro}

Message authentication enables a recipient to verify that a message stems from the claimed sender~\cite{1996_bellare_keying}.
The most prominent and widely-used approaches for such verification are message authentication codes~(MACs), which are used to add authentication tags~(short: \textit{tags}) to each message~\cite{2009_goldreich_cryptography}.
Naturally, these tags expand packets: 
To achieve the minimum security level of \unit[128]{bit} recommended by NIST~\cite{2020_nist}, \unit[16]{bytes} have to be added to a message.
In resource-constrained environments, \eg industrial control systems~\cite{2014_frotzscher_requirements, 2018_hiller_antedated, 2020_serror_qwin}with messages as small as a single byte~\cite{2012_galloway_ics}, a large fraction of a packet is thus consumed by this tag.
In these environments, the expansion of packets is problematic because of 
\begin{enumerate*}[label=(\roman*)]
	\item payloads that are only a few byte long~\cite{2020_kneib_easi},
	\item bandwidth limitations~\cite{2019_glebke_data},
	\item energy restrictions~\cite{2008_cheng_energy}, and
	\item reliability requirements~\cite{2020_serror_qwin}.
\end{enumerate*}

To overcome this obstacle, early work proposed to reduce tag sizes by truncating tags~\cite{1995_preneel_mdx, 2017_wang_survey, 2011_schweppe_car2x, 2008_szilagy_flexible} or aggregating tags of multiple messages~\cite{1995_bellare_xormac, 1997_gennaro_sign,2008_katz_aggregated-mac, 2010_eikemeier_history, 2012_kolesnikov_mac}.
These reductions, however come at the cost of reduced security or intolerable delays.
\emph{Progressive} message authentication codes~(ProMACs)~\cite{2017_schmandt_minimac, 2020_armknecht_promac, 2020_li_cumac, 2021_li_cumacs} address these drawbacks by immediately providing reduced integrity protection upon reception of a message, which is then progressively reinforced by subsequent messages, eventually achieving ``full'' security.
Often, the initial protection suffices to optimistically process messages since the system could recover from the unlikely scenario of a retrospectively~(within seconds) detected attack~\cite{2008_szilagy_flexible, 2010_szilagyi_low, 2017_castellanos_retrofitting2,2019_mashima_optimization}. %

While ProMACs promise to provide strong and low-latency integrity protection even over wireless and other lossy channels, we show that this is not the case for current ProMAC schemes:
Random transmission failures or a network-level adversary, \ie an attacker with the ability to drop or alter packets, can deliberately remove the integrity protection from a complete sequence of messages by interfering with only two carefully chosen packets.
Thus far, ProMACs do not consider the collateral damage of packet loss, \ie the impact of lost packets on the verifiability of surrounding tags and attacks emerging from it. 
To address resulting vulnerabilities, it is imperative to decouple the dependency of tags on a sequence of directly subsequent messages to prevent attackers from voiding the integrity of messages by interfering with a few selected packets.

\textbf{Contributions.}
To enable the \emph{secure} utilization of progressive message authentication, we make the following contributions:
\begin{itemize}[noitemsep,topsep=0pt,leftmargin=9pt]
	\item 
	We show that current ProMAC schemes are prone to a \emph{sandwich attack}, where an adversary selectively attacks the two messages surrounding a message sequence to remove integrity protection of the complete sequence~(Section\,\ref{sec:attack}).
	
	\item To increase ProMACs' resilience to such attacks as well as transmission failures, we propose \emph{\dname{}}, our generic solution to decouple the dependency of tags on a direct message sequence.
	\dname{} builds on the properties of Golomb Rulers to achieve optimal verification delays for predefined security guarantees, which it pairs with randomness and immediate protection bits to achieve \textit{randomized and resilient dependency distribution}~(Section\,\ref{sec:design}).
	
	\item We design and implement \emph{\name}, a ProMAC scheme for resource-constrained devices that realizes \dname{}'s mitigations using efficient XOR operations.
	Our evaluation shows that \name{} effectively protects against network-level interference while operating as resource-conscious as current ProMACs~(Section\,\ref{sec:construction}).
\end{itemize}

\textbf{Availability Statement.}
The source code underlying this paper is available at: \url{https://github.com/fkie-cad/spmac}

\section{Progressive Authentication}
\label{sec:background}

A major challenge of secure communication in resource-constrained scenarios stems from the overhead of integrity protection:
Even the tiniest message requires a tag of several bytes~(\eg \unit[16]{bytes} for 128-bit security), thus significantly increasing messages sizes.
In this section, we 
motivate the core idea of ProMACs, a recent proposal to address this issue~(Section\,\ref{sec:applications}), together with a motivating example~(Section\,\ref{sec:motivating-example}).
Afterward, we formally introduce ProMACs~(Section\,\ref{sec:formal-definition}) and present three practical implementations~(Section\,\ref{sec:schemes}).

\subsection{Core Idea and Benefits of ProMACs}
\label{sec:applications}

Traditional MACs~(\eg HMAC~\cite{2006_bellare_hmac}) occupy large parts of the total payload for short messages.
To partly mitigate this issue, tags can be truncated at the cost of reduced security~\cite{1995_preneel_mdx}.
To provide short tags with strong security guarantees, the core idea behind ProMACs is to partially offload integrity protection into the future. 
Therefore, each message is initially only protected with a reduced security level, similar to truncated MACs, but subsequent messages quickly increase this protection to an adequate level~(\eg \unit[128]{bit}).
Since tags aggregate the protection of multiple messages at once, ProMACs realize short tags, while enabling passive resynchronization if packets get lost.
This passive resynchronizability is in stark contrast to previous proposals for short tags, such as aggregated MACs, which jointly authenticate multiple messages with a single tag, and stateful MACs, which continuously reinforce the integrity of all previously sent messages but cannot cope with packet loss.

Consequently, ProMACs are proposed for various (wireless) scenarios such as vehicular communication~\cite{ 2017_schmandt_minimac, 2020_li_cumac, 2021_li_cumacs, 2020_armknecht_promac,2008_nilsson_efficient}, (industrial) IoT~\cite{2020_li_cumac, 2021_li_cumacs, 2020_armknecht_promac}, drone control~\cite{2020_armknecht_promac, 2017_bachhuber_minimization}, and internal communication within hardware components~(\eg Intel SGX or SoCs)~\cite{2020_armknecht_promac, 2016_gueron_sgx, 2018_moriam_soc}.
To cope with the low latency requirements of those scenarios, ProMACs rely on optimistic security, which (partly) defers security processing into the future and allows a system to continue under the assumption that all traffic is benign.
In the unlikely event that an attack is detected retrospectively, the system recovers from already processed malicious messages.
Optimistic security is especially attractive in isolated networks where attacks are relatively rare and the potential damage in a short time frame is comparable to that of less advanced attacks~(\eg~denial of service)~\cite{2008_szilagy_flexible, 2008_zhang_raise, 2009_szilagyi_flexible, 2010_szilagyi_low, 2012_szilagyi_phd, 2017_castellanos_retrofitting2,2019_mashima_optimization}.

\subsection{A Motivating Example for ProMACs}
\label{sec:motivating-example}
To illustrate how ProMACs' benefits manifest themselves in practice, we consider a comprehensive example from an \ac{ICS}.
In particular, we envision a closed-loop motion controller that reacts to continuously updated sensor readings~\cite{VDE-Position-Paper}.
Especially for moving systems, wireless, and thus unreliable, communication is used to avoid error-prone and expensive cable management~\cite{VDE-Position-Paper}.
While such a controller is resilient to the immediate impact of individual faulty sensor readings, one or multiple maliciously crafted messages can, over time, bring the system into an equipment-damaging or even life-threatening state.
Meanwhile, communication channels between sensors and controllers are often constrained due to \eg a high number of network participants.

While bandwidth constraints prevent the traditional protection of each message with a 16\,byte long tag, using a truncated (\ie less secure) tag hampers the reliable detection of manipulations.
Therefore, an attacker could manipulate messages with long-term impact, \eg a scaling factor for speed adjustments, and thus bring the system into a critical state.
While aggregated MAC schemes could eventually ensure integrity with high confidence, they would give an attacker the opportunity to manipulate multiple messages before any authenticity is verified.
ProMACs mitigate this weakness by providing, albeit reduced, immediate security.
ProMACs thus protect against the immediate impact of manipulated messages, while also protecting against manipulations with long-term impact.

\subsection{Formal Definition of ProMACs}
\label{sec:formal-definition}

To explain how ProMACs realize efficient protection, we formally introduce them based on traditional MACs.

\textbf{Traditional MACs.}
A MAC scheme allows two communication partners to authenticate exchanged messages using a pre-shared secret $k$.
To authenticate a message $m$, the sender uses the tag generation algorithm $\textit{Sig}_k(m)$ to generate the corresponding authenticity tag $t$.
Upon reception of a message, the verification algorithm $\textit{Vrfy}_k(m,t)$ enables the recipient to evaluate whether the received tag is valid. %
This verification is done by computing the tag for the received message $m$ and comparing it to the received tag $t$.
A MAC scheme is considered secure if it is computationally infeasible to generate a ($m$,$t$)-pair that would be accepted by $\textit{Vrfy}_k$, without knowing the secret $k$.
This requirement can, \eg be achieved by using keyed hash functions such as \code{HMAC-SHA256} to compute $t$.

\begin{figure}
	\centering	
	\includegraphics[width=0.9\columnwidth]{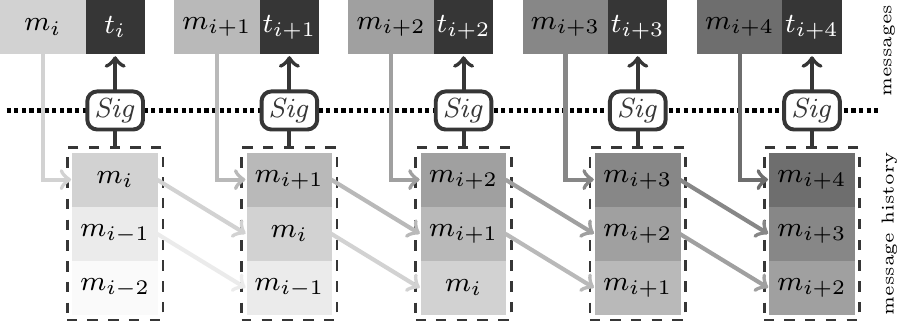}
	\caption{
		ProMACs store a history of recent messages, which is used to derive tags, effectively aggregating integrity protection of multiple messages to reduce tag sizes. 
	}
	\label{fig:promacs}
\end{figure}

\textbf{ProMACs.}
ProMACs extend traditional MACs by additionally giving recent historical messages as input to $\textit{Sig}_k(m_1,\dots, m_n)$ and  $\textit{Vrfy}_k(m_1,\dots, m_n,t)$.
As shown in Figure\,\ref{fig:promacs}, the generation of tags is then based on \emph{multiple} messages.
The tag $t_{i+2}$, for instance, is computed from messages $m_{i+2}$, $m_{i+1}$, and $m_{i}$.
Likewise, the integrity of message $m_{i+2}$ is protected by tags $t_{i+2}$, $t_{i+3}$, and $t_{i+4}$.
Thus, ProMACs protect each message with multiple tags, which means that each tag is only responsible for providing a fraction of the overall targeted security level.
Since each tag aggregates partial integrity protection for multiple messages, progressive integrity protection results in shorter tags. 
Meanwhile, a valid first tag (to the degree it can be verified) is considered sufficient to optimistically process a message, while a recovery mechanism is triggered if an attack is detected within subsequent tags.
In this context, the dependencies $\mathcal{D}$~$(\{0\} \subseteq \mathcal{D} \subset \mathbb{N}_0 )$ describe how the reception of one message influences the authenticity of surrounding messages.
We say that a ProMAC instance has the dependencies $\mathcal{D}$, if the generation and verification of tag $t_i$ require knowledge of $\{ m_{i-d} | d\in\mathcal{D} \}$.
Consequently, a message $m_i$ blends into all tags $\{ t_{i+d} | d\in\mathcal{D} \}$ and a tag $t_i$ protects the integrity of all messages $\{ m_{i-d} | d\in\mathcal{D} \}$.

\subsection{Existing ProMAC Schemes}
\label{sec:schemes}

Three distinct approaches realize the theoretical notion of ProMACs:
\textit{Whips}~\cite{2020_armknecht_promac}, CuMAC~\cite{2020_li_cumac, 2021_li_cumacs}, and \textit{Mini-MAC}~\cite{2017_schmandt_minimac}.
Our discussion of these approaches focuses on their selection of identical dependencies $\mathcal{D}$, as those describe how the failure to receive one message influences the verifiability of neighboring tags. 

\textbf{Whips \emph{(CCS'20)}.}
Whips~\cite{2020_armknecht_promac} was proposed alongside the formal introduction of ProMACs.
It provides a fixed security level of \unit[128]{bit} with a constant memory overhead per message stream.
To this end, Whips tracks the message history via an internal state $s$, used to derive tags $t$, that is composed of a counter $c$ (for replay protection) and a fixed number $n$ of substates $\tilde{s}$.
The number of substates is inversely proportional to the targeted tag lengths, such that if smaller tags are used, a message's integrity is protected by more tags.
Each substate $\tilde{s}_i$ corresponds to exactly one message and is computed as $\tilde{s}_i=trunc($\code{HMAC}$_k(m_i))$.
The size of $\tilde{s}$ depends on the targeted security level~(\eg $\tilde{s}$ has to be at least \unit[32]{byte} long for 128-bit~security~\cite{2020_armknecht_promac}). 
To generate a new tag $t_i$ for a message $m_i$, the state $s_i$ is first updated by
\begin{enumerate*}[label=(\roman*)] 
	\item incrementing the counter $c$, 
	\item appending the substate $\tilde{s}_i$ to $s$, and 
	\item removing the substate $\tilde{s}_{i-n}$ from $s$.
\end{enumerate*}
The tag $t_i$ for message $m_i$ is then computed as $trunc($\code{HMAC}$_k(s_i))$.
Since a tag thus depends on the last $n-1$ messages, Whips relies on sliding window-based dependencies, \ie $\mathcal{D} = \{0, \dots, n-1\}$.

\textbf{CuMAC \emph{(CNS'20 \& IoT-J'21)}.}
Simultaneous to the formalization of ProMACs~\cite{2020_armknecht_promac}, CuMAC~\cite{2020_li_cumac} proposed the similar concept of cumulative message authentication codes.
In CuMAC, first, a traditional MAC $\sigma$ is computed from a counter $c$ and a message $m$.
Then, $\sigma$ is split into $n$ fragments, \ie $\sigma = \sigma^0||\dots||\sigma^{n-1}$.
Finally, the tag $t_i$ for message $m_i$ is computed by aggregating fragments of the MAC $\sigma$ for the $n$ past messages using XOR (one distinct fragment per message). 
More precisely, $t_i=\sigma^{0}_i \oplus \sigma^{1}_{i-1} \oplus \dots \oplus  \sigma^{n-1}_{i-n}$.
Thus, CuMAC also relies on a sliding window of the $n$ most recent messages ($\mathcal{D} = \{0, \dots, n-1\}$).
To further improve CuMAC, CuMAC/S~\cite{2021_li_cumacs} tries to predict future messages and pre-authenticates these messages to achieve immediate full authentication upon message reception.
However, this does not change its dependencies $\mathcal{D}$.

\textbf{Mini-MAC \emph{(Veh.\,Comm.'17)}.}
Mini-MAC~\cite{2017_schmandt_minimac} re-authenticates CAN bus messages within subsequent messages to address the problem of insufficient payload size. 
Although originally not designed as generally applicable, retrospectively Mini-MAC can be interpreted as ProMAC scheme if we ignore optional extensions.
Mini-MAC derives a tag $t_i$ (for message $m_i$) from a sliding window of the $n$ most recent messages ($\mathcal{D} = \{0, \dots, n-1\}$) and a counter $c$ (for replay protection):
$t_i=\textit{trunc}($\code{HMAC}$_k(c\|m_{i-(n-1)}\|\dots\|m_i))$.
The size of the sliding window $n$ is not fixed.
A larger $n$ results in higher eventual security, but also requires more computations and increases the impact of transmission failures.
Additionally, $t_i$ is truncated to the space remaining in the given packet.
Consequently, Mini-MAC provides integrity protection in a best-effort manner.

\section{Security Consideration for ProMACs}
\label{sec:attack}

Security of ProMACs so far centered around an attacker with the same goal and means as for traditional MACs, \ie attacking individual packets by guessing keys or forging tags~\cite{2017_schmandt_minimac,2020_armknecht_promac, 2020_li_cumac, 2021_li_cumacs}.
In this setting, ProMACs provide at least the same security as traditional MACs:
For the latest message, the security of ProMACs is allegedly identical to traditional (truncated) MACs and becomes stronger with subsequent packets~\cite{2020_armknecht_promac}. 
However, these security considerations ignore the impact of \textit{dropped} packets, which influence the verifiability of neighboring tags. 
Hence, we extend ProMACs' threat model~(Section\,\ref{sec:attack:threat-model}) and show that this leads to novel attacks~(Section\,\ref{sec:attack:attack}) that severely limit the applicability of current ProMACs.

\subsection{Extended Threat Model for ProMACs}
\label{sec:attack:threat-model}

To accommodate for ProMACs spreading authenticity over multiple packets, some of which may be lost due to a lossy~(\eg wireless) channel, we extend the original threat model of ProMACs~\cite{2020_armknecht_promac} in two ways.
Firstly, we extend the attackers' capability beyond simply observing and querying message-tag pairs by giving them the additional capability of inducing and reacting to transmission failures.
Secondly, we alter the attackers' goals to include not only the forging of valid tags for a previously unseen message but also the disruption of the communication channel by \eg amplifying a DoS attack by abusing the characteristics of ProMACs.

\subsection{Sandwich Attack Against Current ProMACs}
\label{sec:attack:attack}

\begin{figure}
	\centering	
	\includegraphics[width=0.9\columnwidth]{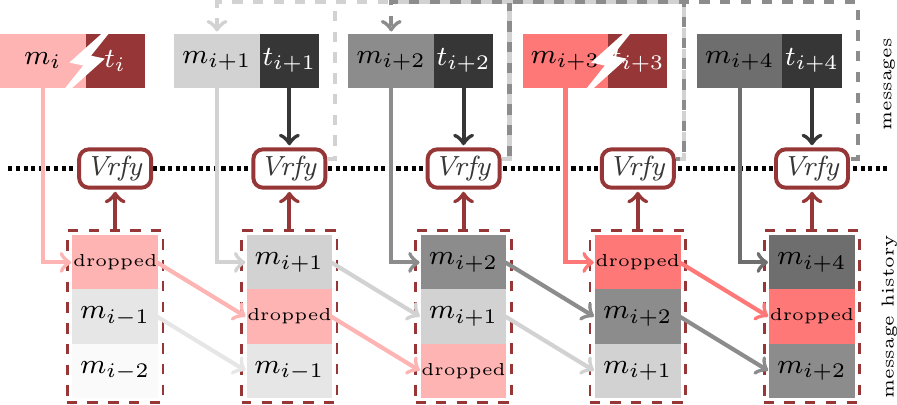}
	\caption{ 
		Sliding window-based dependencies, as used by all current state-of-the-art ProMACs schemes, allow an attacker to remove integrity protection from multiple messages by sandwiching them between dropped packets.		
	}
	\label{fig:promacs_attack}
\end{figure}

So far, ProMACs did not consider the effects of transmission failures, either caused by a lossy channel or active interference, in their (formal) security proofs~\cite{2020_armknecht_promac, 2021_li_cumacs}.
The \emph{sandwich attack} against ProMACs presented in this paper leverages exactly this attack vector:
If two transmission failures are less than the tracked message history apart, all messages ``sandwiched'' between these failures remain unauthenticable.
Selective jamming to induce these failures is, however, hardly distinguishable from random packet loss, such that these attacks are hard to detect as only a small number of dropped packets can have detrimental consequences.

Figure\,\ref{fig:promacs_attack} explains the root cause of this attack using an example with a message history of length $n=3$ (chosen short for illustration).
We assume that message $m_i$ is not received, either through a transmission failure or jamming. 
Then, because of the dependencies $\mathcal{D} = \{ k | 0 \leq k < n \}$ of current ProMACs, all future tags $t_{i+k}$ $(k < n)$ cannot be verified. 
In itself, this is not a serious problem, as eventually, messages $m_{i+k}$ ($k<n$) will still receive (reduced) integrity protection through the tags $t_{i+k+j}$ ($n-k+i$<$j$<n).
Therefore, in the primarily envisioned scenarios for ProMACs~(\cf~Section~\ref{sec:applications}), $m_{i+1}$ and $m_{i+2}$ would be processed optimistically under the assumption that messages stemming from an attacker would be detected before any real damage could occur.
However, if another message $m_j$ $(i+1<j\leq n)$ is also not received (by chance or triggered by interference), then all tags $\{ t_{i}, t_{i+1}, \dots, t_{j+n-1}, t_{j+n} \}$ cannot be verified. 
Consequently, all messages $m_l$ $(i<l<j)$, sent in between $m_i$ and $m_j$, cannot be authenticated, as their integrity protection relies on the tags $\{ t_{l+k} | k \in \mathcal{D} \}  \subset \{ t_{i}, t_{i+1}, \dots, t_{j+n-1}, t_{j+n} \}$.
Already for our selected short message history, the authenticity of $m_{i+1}$ and $m_{i+2}$ cannot be verified despite being received correctly.

Overall, the prospects of ProMACs to bandwidth-efficiently protect lossy communication are highly desirable.
However, their susceptibility to network-level disturbances limits ProMACs' deployability.
Looking at the practical consequences of the sandwich attack in Appendix~\ref{app:attack-in-practice}, we further see that the attack is more impactful for shorter tags due to their larger sliding windows.
Consequently, those scenarios that benefit most from ProMACs' bandwidth savings are the most vulnerable to the presented sandwich attack.

\section{R2-D2: A Basis for Secure ProMACs}
\label{sec:design}

\begin{figure}
	\centering	
	\includegraphics[width=\columnwidth]{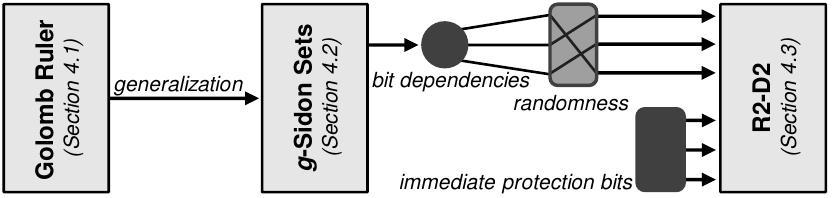}
	\caption{\dname{} combines different theoretical building blocks to realize randomized and resilient dependency distribution and to thwart sandwich attacks against ProMACs.}
	\label{fig:r2d2_overview}
	\vspace*{-.9em}
\end{figure}

The weakness of current ProMAC schemes stems from an inherent design choice:
By spreading a message's integrity protection over \emph{consecutive messages}, these schemes become susceptible to packet loss, where 
the (malicious) interference on a few packets invalidates the authenticity for multiple messages~(\cf\ Section\,\ref{sec:attack}).
To eliminate this attack vector, it is necessary to \emph{interleave message dependencies} 
to better distribute the effects of dropped packets 
to prevent individually lost packets from reducing the protection of targeted messages to insecure levels.
However, while the general idea of interleaving message dependencies seems promising, it requires finding the right trade-off between delay for full integrity protection and resilience to dropped packets.
To achieve this goal, we propose our Randomized and Resilient Dependency Distribution~(\dname) as a foundation for ProMACs that are resilient to network-level interference. 
As shown in Figure\,\ref{fig:r2d2_overview}, \dname is based on 
\emph{optimally} interleaved dependencies~%
(Section\,\ref{sec:golomb-ruler}) and a generalization of this concept to achieve \textit{parameterized security guarantees}~(Section\,\ref{sec:param-sec-guarantees}).
\dname enhances this foundation through \textit{bit dependencies}, \textit{randomization}, and \textit{immediate protection bits}~(Section\,\ref{sec:design-bit-dependencies}).

\subsection{Golomb Ruler-based Dependencies}
\label{sec:golomb-ruler}

Existing ProMAC schemes rely on a sliding window for their dependency distribution, where each tag's computation requires knowledge of the last $n$ consecutive packets, \ie $\mathcal{D} = \{ 0,\dots,n \}$.
However, exactly this property is abused by the sandwich attack to render the integrity protection of targeted messages void.
To mitigate this weakness, ProMACs have to interleave dependencies such that the effects of dropped packets are cushioned by a large set of tags.
To achieve this goal, we propose to use \emph{Golomb Rulers}~\cite{1932_sidon_satz,1953_babcock_gr}, which are used, \eg in radio astronomy to determine optimal antenna placements~\cite{2017_richard_antenna}, to minimize overlap between tags of different messages.
Intuitively, a Golomb Ruler is a set of integer marks on a discrete ruler, placed such that the distance between any pair of marks is unique.
Formally, a set $S$~($\{0\} \subseteq S \subset \mathbb{N}_0$) is a Golomb Ruler iff $\forall s_1,s_2,s_3,s_4 \in S$ with $ s_1 \neq s_2$ and $s_3 \neq s_4$ it holds that $ s_1 - s_2 = s_3 - s_4 \iff s_1 = s_3 $ and $ s_2 = s_4$. 
The length of a Golomb Ruler is defined as the value of its largest element.

\begin{figure*}
	\begin{subfigure}{\columnwidth}
		\centering
		\includegraphics[width=\textwidth]{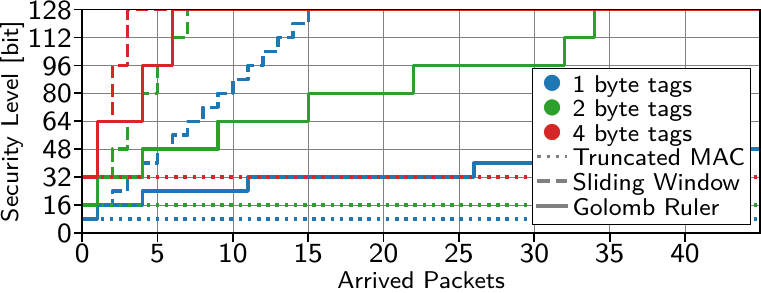}
		\caption{Golomb Ruler-based dependencies require the reception of more messages to achieve full 128-bit~security than sliding window approaches, and thus increase the delay of integrity protection in a message stream, particularly for small tags~($\leq$\unit[2]{byte}).}
		\label{fig:ogr_tags}
	\end{subfigure}
	\hfill
	\begin{subfigure}{\columnwidth}
		\centering
		\includegraphics[width=\textwidth]{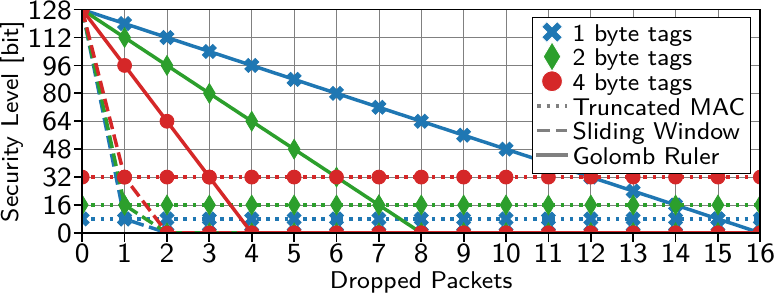}
		\caption{Dropped packets have an obvious impact on the progressive nature of ProMACs. However, the resiliency to dropped messages is significantly increased by Golomb Ruler-based ProMACs compared to current state-of-the-art ProMACs.}
		\label{fig:ogr_drop}
	\end{subfigure}
	\vspace*{-.6em}
	\caption{Golomb Ruler-based ProMACs~(in comparison to sliding window-based ProMACs and truncated MACs) protect against network-level attacks, albeit with a stiff tag length-dependent trade-off between speed and security.}
	\label{fig:ogr}
\end{figure*}

Golomb Rulers provide the theoretical foundation to realize message dependencies that minimize the overlap between tags of neighboring messages.
As the distance between two tags protecting a specific message is unique, Golomb Rulers guarantee that a message's security level is reduced by at most the integrity protection provided by one tag for any dropped message (see~proof in Appendix~\ref{app:proof-gr}).
Exemplarily, using the Golomb Ruler $\{0,1,4,6\}$ of length 6 ensures that any dropped message only invalidates at most one tag protecting the integrity of any other message,
while providing full security guarantee after 6 subsequent messages have been received.
Using the shortest Golomb Rulers for a given number of elements, \ie an optimal Golomb Ruler, as in the previous example, thus provides the mentioned security guarantees within the shortest possible delay until full authenticity is reached.

We now have to investigate what Golomb Ruler-based dependencies mean in general for the security of ProMACs, based on two core metrics: \begin{enumerate*}[label=(\roman*)]
	\item the \textit{delay} to reach full protection, and
	\item the \textit{resilience} to targeted packet dropping.
\end{enumerate*}
Here, delay is expressed as transmitted packets, as actual time depends on the communication pattern of the underlying application.
Resilience is expressed by the minimum obtainable bit security of any message for a given number of lost packets.
To express resilience in terms of bit security, we assume that a $n$-bit tag provides exactly $n$\,bits of security, \ie 128-bit security is realized by appending a 16-byte tag to a message.

In Figure\,\ref{fig:ogr}, we show the delay and resilience of Golomb Rulers-based ProMACs by comparing them to truncated MACs and sliding windows-based ProMACs for tags that are 1, 2, and \unit[4]{byte} long. 
Figure\,\ref{fig:ogr_tags} shows how the protection develops over time in absence of an attacker.
We observe a fast increase in security for sliding window-based ProMACs and no variation in the provided security over time for truncated MACs.
Using optimal Golomb Rulers as dependencies, the delay until full security is reached is acceptable for tag sizes of 4 and \unit[2]{byte}, whereas 1-byte tags only reach full security after receiving 177 additional messages. 

In contrast, Figure\,\ref{fig:ogr_drop} shows the resilience of different schemes against network-level attacks.
Here, the susceptibility to the sandwich attack of sliding window-based ProMACs can be seen again, as the provided protection of a message can be rendered void with just 2 dropped packets.
Truncated MACs are, as expected, not susceptible to network-level attacks.
When it comes to the resilience of Golomb Ruler-based ProMACs, we see an inverted behavior as in Figure\,\ref{fig:ogr_tags}.
1 and \unit[2]{byte} long tags provide significant resilience to network-level attacks, remaining well over the security level of truncated MAC, even if a high fraction of the relevant packets are dropped.
However, longer tags remain susceptible to variants of the sandwich attack, \ie the integrity of message protected by  \unit[4]{byte} long ProMACs can be attacked by dropping 4 targeted messages.

Golomb Ruler-based dependencies provably minimize the delay until full security while ensuring that a dropped packet impacts at most one tag protecting any other message.
However, the resulting inflexible trade-off between achievable delay and resilience might not match the requirements of specific use cases.
We thus discuss how \dname addresses this challenge via generalized Golomb Rulers.

\begin{figure*}
	\begin{subfigure}{\columnwidth}
		\centering
		\includegraphics[width=\textwidth]{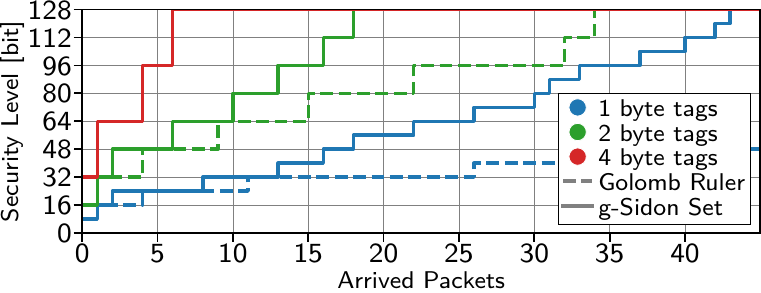}
		\caption{Using $g$-Sidon Sets instead of Golomb Rulers can significantly reduce the delay until the full security level is reached.}
		\label{fig:sidon_tags}
	\end{subfigure}
	\hfill
	\begin{subfigure}{\columnwidth}
		\centering
		\includegraphics[width=\textwidth]{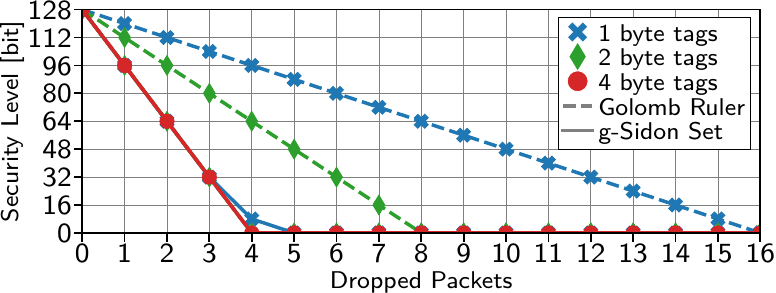}
		\caption{ProMACs based on $g$-Sidon Sets enable parameterized tag length-independent security loss resulting from dropped packets.}
		\label{fig:sidon_drop}
	\end{subfigure}
	\vspace*{-.6em}
	\caption{$g$-Sidon Sets, in contrast to Golomb Rulers, enable to control the security loss per dropped packet~(here:\,\unit[32]{bits}).}
	\label{fig:sidon}
\end{figure*}

\subsection{Tag Length-independent Security Levels}
\label{sec:param-sec-guarantees}

Message dependencies with minimal overlap based on Golomb Rulers directly couple the security loss from a dropped packet to the length~(and thus bit security) of individual tags~(\cf~Section\,\ref{sec:golomb-ruler}).
Thus, while providing optimal dependencies w.r.t.\ the number of unverifiable tags in any given message, these dependencies may result in unacceptable verification delays 
for certain scenarios.
To resolve this stiff trade-off, we propose \emph{tag length-independent security levels} for providing message dependencies that enable a parametrization of the maximum security loss per dropped packet. 

This parametrization enables to define tags of different sizes that each provide similar resilience to network-level attacks.
The core idea to achieve tag length-independent security levels in \dname{} is to give control over how many tags, protecting the integrity of a single message, become at most unverifiable through a dropped packet.
To realize this idea, we use \textit{$g$-Sidon Sets}~\cite{1932_sidon_satz}, a generalization of Golomb Rulers.
Intuitively, a $g$-Sidon Set is a set of integer marks on a discrete ruler, which are placed such that the distance between any pair of two marks occurs at most $g$ times.
Formally defined, a set $S \subset \mathbb{N}_0$ is a $g$-Sidon Set iff any pairwise difference between elements occurs at most $g$ times, \ie $S$ is a $g$-Sidon Sets iff there exist at most $g$ distinct pairs~($s_0\in S$, $s_1\in S$) with $s_0<s_1$ such that $s_0-s_1=k$, for all $k\in \mathbb{Z^*}$.
Using $g$-Sidon Sets as ProMAC dependencies $\mathcal{D}$ guarantees that any message's security level is reduced by at most the integrity protection provided by $g$ tags for any dropped message~(see proof in Appendix~\ref{app:proof-ss}). 
Thus, Golomb Rulers are $1$-Sidon Sets, as any difference between elements in a Golomb Ruler is unique.
Overall, these provable and parameterized level of protection against network-level attacks is possible iff $g$-Sidon Sets are used as dependencies, while their optimality guarantees that full security is achieved in the fastest possible way.

For ProMACs, $g$-Sidon Sets thus promise to efficiently parameterize the maximum security loss for dropped packets in terms of bit security to decouple this property from the tag length and gain control over verification delays.
To verify this claim, we compare verification delays and resilience to network-level attacks of Golomb Ruler-based dependencies to those based on $g$-Sidon Sets in 
Figure\,\ref{fig:sidon}.
We choose $g$ such that a dropped message induces at most a 32-bit security loss, independent of the underlying tag size~(\ie~$g=1$ for 4-byte tags, $g=2$ for 2-byte tags, and $g=4$ for 1-byte tags).
Figure\,\ref{fig:sidon_tags} shows the improvements to verification delays based on parameterization.
By allowing dependencies to overlap twice, we can nearly half the verification delay of 2-byte tags, and the verification delay of 1-byte tags can be reduced from 177 to 43.
However, this speedup also reduces the resilience to attacks, as can be seen in Figure\,\ref{fig:sidon_drop}.
Here, we show the advantage of $g$-Sidon Sets-based dependencies since all parameterizations lose security to a similar extent with the number of dropped packets, but this loss is bounded by the maximal security loss of \unit[32]{bits} per dropped packet.
Thus, a variation of the sandwich attack would require the targeted dropping of at least four packets to remove all authenticity.

Dependencies based on $g$-Sidon Sets thus achieve tag length-independent security levels and allow a flexible parameterization of the trade-off between verification delays and resilience to packet loss.
We observe that full security can be provided significantly faster for smaller tags optimal $g$-Sidon Sets than by Golomb Rulers-based dependencies, which is counteracted by a reduction in resilience to packet loss.
Additionally, using optimal $g$-Sidon Sets lets an attacker know the optimal strategy to remove integrity protection from a targeted message, whereas it would be advantageous to hide this strategy.
In the following, we see how \dname addresses these remaining weaknesses through bit dependencies, hiding of the optimal attack strategy, and immediate protection bits.

\subsection{Secure Dependencies through R2-D2}
\label{sec:design-bit-dependencies}

Both, message dependencies with minimal overlap~(Section\,\ref{sec:golomb-ruler}) and its more flexible generalization for tag length-independent security levels~(Section\,\ref{sec:param-sec-guarantees}), realize optimal and thus \emph{deterministic} dependencies for their respective parametrizations.
Consequently, while these improved dependencies considerably increase the number of necessary 
packet drops to disable integrity protection,
an attack can still leverage this determinism to derive \emph{which} packets to drop.

\begin{figure*}
	\begin{subfigure}{\columnwidth}
		\centering
		\includegraphics[width=\textwidth]{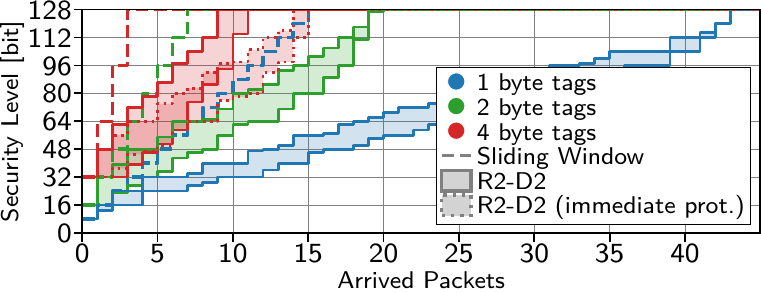}
		\caption{R2-D2 introduces randomness into its bit dependencies without significantly increasing the delay for full integrity protection. }
		\label{fig:r2d2_tags}
	\end{subfigure}
	\hfill
	\begin{subfigure}{\columnwidth}
		\centering
		\includegraphics[width=\textwidth]{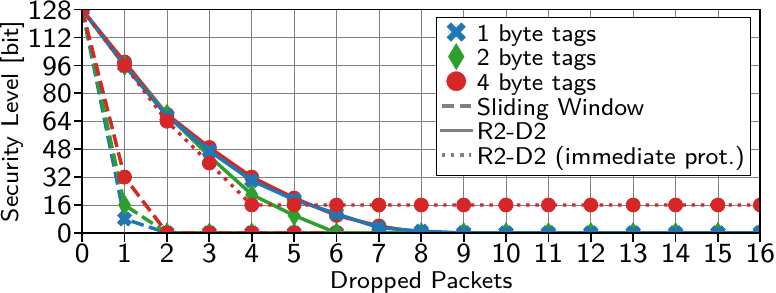}
		\caption{
			R2-D2 protects against sandwich attacks, even if the attacker learns the randomized bit dependencies~(worst-case assumption).
		}
		\label{fig:r2d2_drop}
	\end{subfigure}
	\vspace*{-.6em}
	\caption{ Even in the worst-case~(\ie the attacker somehow learns the secret bit dependencies), R2-D2 offers high resilience to network-level attacks without significant compromises in terms of verification delay. }
	\label{fig:r2d2}
\end{figure*}

\dname{} addresses this issue by \emph{randomizing} dependencies to hide which messages have to be dropped.
Further enhancing this approach, \dname{} introduces \emph{bit dependencies}, \ie each bit of a tag protects a different message set.
Each \dname{} instance
is initialized with a \emph{pseudorandom} set of dependencies $\mathfrak{D}=\{\mathcal{D}_0, \mathcal{D}_1, \dots\}$, where the number of dependencies equals the tag length, \ie $|\mathfrak{D}|=|t|$.
Each dependency $\mathcal{D}_i$ is a $g$-Sidon Set, where its order, \ie number of elements, depends on the tag length $|t|$ such that the total number of dependencies equals the targeted security level, \eg $\sum_{0\leq i < |t|} |\mathcal{D}_i| = 128$.
The parametrization of $g$ follows from the tolerable security loss~(\cf\ Section\,\ref{sec:param-sec-guarantees}).
The set of dependencies $\mathfrak{D}$ is pseudorandomly sampled (using a shared key between sender and receiver) from the $n$ precomputed most optimal $g$-Sidon Sets. 

Instead of selecting one of $n$ potential dependencies, the number of potential distributions increases to $\binom{n}{|t|}$ through randomized bit dependencies.
This increased variety enables strong resilience with a relatively short $n$, \eg 64, meaning that verification delay remains low and that even constrained devices can store the set of potential bit dependencies.
Additionally, by using dependencies of different orders, \dname{} can achieve a specific security level, \eg 128\,bits, even if the targeted tag length does not divide the security level, since the achieved security level amounts to $\sum_{0\leq i < |t|} |\mathcal{D}_i|$\,bits.

As an additional benefit, bit dependencies enable \textit{immediate protection bits}.
Those only depend on the current message~($\mathcal{D}_i = \{0\}$) and thus, like truncated MACs, are resilient to network-level interference.
This ensures that, no matter how many packets are dropped by an attacker, the protection of a received message is never completely removed.
To still reach the targeted bit security level when using immediate protection bits, the order of the remaining dependencies $\mathcal{D}_i\in\mathfrak{D}$ has to be increased accordingly.

The concepts of randomized bit dependencies and immediate protection bits promise to increase the resilience to network-level attackers without significant impacts on the verification delay.
To verify this claim, we again compare the verification delay and resilience of \dname to sliding window-based dependencies, the current state-of-the-art in Figure\,\ref{fig:r2d2}.
In addition to the 1, 2, and \unit[4]{byte} tags with randomized bit dependencies, we also consider the case where half of a 4-byte tag is reserved for immediate protection bits.
As before, all tag lengths are parameterized to allow a maximal security loss of \unit[32]{bits} per dropped packet (further parameters are presented and discussed in Appenix~\ref{app:more-results}).
All dependencies are selected from the $64$ shortest ones for a given parametrization, \ie $n=64$.

In Figure\,\ref{fig:r2d2_tags}, the verification delay of \dname{} constitutes an area between the minimum and maximum delay depending on which dependencies are randomly selected.
Overall, we observe similar delays as in Figure\,\ref{fig:sidon_tags}.
Additionally, in Figure\,\ref{fig:r2d2_drop}, we show the worst-case resilience of \dname against a network-level attacker.
Therefore, we assume that 
\begin{enumerate*}[label=(\roman*)]
	\item the attacker knows the selected bit dependencies, and
	\item that the most vulnerable dependencies are selected for the most efficient variation of the sandwich attack.
\end{enumerate*}
We observe that the resilience of \dname increases even in this worst-case scenario through the introduction of bit dependencies, while in reality, the attack would require to drop even more packets as he cannot be sure which packets need to be dropped to execute a sandwich attack.
In practice, this protection against targeted packet drops is even higher, since the randomly selected dependencies remain secret.

Considering the introduction of immediate protection bits to 4-byte tags, we see in Figure\,\ref{fig:r2d2_tags} that this addition results in a slight increase in the delay until full protection is achieved.
The reason for this additional delay is that to still target 128-bit security, higher-order dependencies have to be chosen because fewer bits are available for progressive authentication.
Looking back at Figure\,\ref{fig:r2d2_drop}, we see that this additional delay creates baseline protection that is not susceptible to network-level attacks, no matter how many packets an attacker can drop, similar to truncated MACs~(\cf~Figure\,\ref{fig:ogr_drop}).

\subsection{Security Properties of R2-D2}
\label{sec:security-properties}

Overall, \dname defines parametrizable dependencies which increase the resilience of ProMACs to packet drops, either through a bad channel or malicious activities~(\ie sandwich attacks and variations thereof).
In addition to inheriting the security guarantees of g-Sidon Set-based dependencies, \ie parameterizable bounds for the maximal security loss per dropped packet, \dname additionally hides the optimal attack strategy for an attacker.
A big strength of \dname is its flexibility: ProMACs can be adapted to different use cases by choosing trade-offs between the maximal security loss per dropped packet, tag lengths, and acceptable delays until full authenticity can be provided.
Finally, \dname achieves provably \emph{optimal} authentication delays for given security parameters, such that we can understand
where ProMACs may not be applicable.

\textbf{Practical Authentication Delays.} 
As ProMACs provide only reduced initial security, messages are processed optimistically and may retrospectively be detected as malicious.
Related work on optimistic security~\cite{2010_szilagyi_low, 2012_szilagyi_phd, 2008_nilsson_efficient,2017_castellanos_retrofitting2,2009_szilagyi_flexible} puts resulting delays into perspective to understand where ProMACs are applicable.
For intra-vehicular communication, Szilagyi~\etal, \eg demonstrate that authentication delays of up to 100 messages are acceptable even for throttling control due to high sampling rates and physical inertia~\cite{2010_szilagyi_low, 2012_szilagyi_phd}.
Similarly, Nilsson~\etal conclude that delays of up to 16~messages are easy to recover from~\cite{2008_nilsson_efficient}. 
For \acp{ICS}, Castellanos~\etal show that the optimistic processing of over 100~messages does not significantly impact the systems' state under attack~\cite{2017_castellanos_retrofitting2} and Szilagyi~\etal argue that many \acp{ICS} can handle malicious packets if detected within 30~messages~\cite{2009_szilagyi_flexible}.
While \dname can be parameterized to individual needs, 2 and \unit[4]{byte} long tags~(latter with 16 immediate protection bits) with a maximum security loss of 32\,bit are attractive in many real-world deployments: 
Packets are adequately protected against network-level attacks and reach full security within acceptable delays.
However, \dname's optimal delays show that shorter tags or strong security guarantees with ProMACs, in general, can only be provided if longer delays are acceptable.
\textbf{ProMACs on High-Error Rate Channels.}
\dname lowers the number of ProMAC-protected messages with unverifiable authenticity due to normal packet loss.
Yet, we still measure unverifiable authenticity for around 10\% of messages~(down from 73\,\% for sliding window-based dependencies as seen in Appendix\,\ref{app:attack-in-practice}) 
for 1-byte tags on a channel with a packet loss of 9.1\%.
Thus, for adequate security, the use of \dname's immediate security bits and slightly longer tags are crucial to realize secure ProMACs for such channels.
Meanwhile, for the same 1-byte tags, we did not observe a single message that lost all authenticity over tens of millions of transmitted packets with a packet loss of 0.9\%.
Thus, for higher reliability channels, \dname makes it unlikely that a processed message has to be reverted because it could not be retroactively authenticated without malicious interference.

\section{SP-MAC: A Secure ProMAC Scheme}
\label{sec:construction}

\dname{} provides the necessary building blocks for efficient ProMACs schemes that are resilient to network-level interference.
One of its core features to achieve this protection is a shift from message dependencies to bit dependencies.
As this shift fundamentally changes the interplay between tag aggregation and cryptographic operations, current ProMAC schemes (Whips~\cite{2020_armknecht_promac}, CuMAC~\cite{2020_li_cumac, 2021_li_cumacs}, and Mini-MAC~\cite{2017_schmandt_minimac}) cannot easily be retrofitted with \dname{}'s improvements.
Therefore, we propose a novel ProMAC scheme for \emph{staggered} progressive message authentication codes~(\name{}) that leverages traditional and secure MACs~(\eg \code{HMAC-SHA256}~\cite{2006_bellare_hmac}) and aggregates these based on \dname{}'s secure dependency distribution using efficient XOR operations.
As a result, \name{} does not only provide built-in protection against network-level attacks but even achieves this as resource-consciously as existing ProMAC schemes.
In the following, we introduce \name{}~(Section\,\ref{sec:construction:design}), discuss its security~(Section\,\ref{sec:construction:security}), and evaluate its performance~(Section\,\ref{sec:construction:eval}).

\subsection{Staggered Progressive MACs}
\label{sec:construction:design}

\begin{figure}
	\centering	
	\includegraphics[width=\columnwidth]{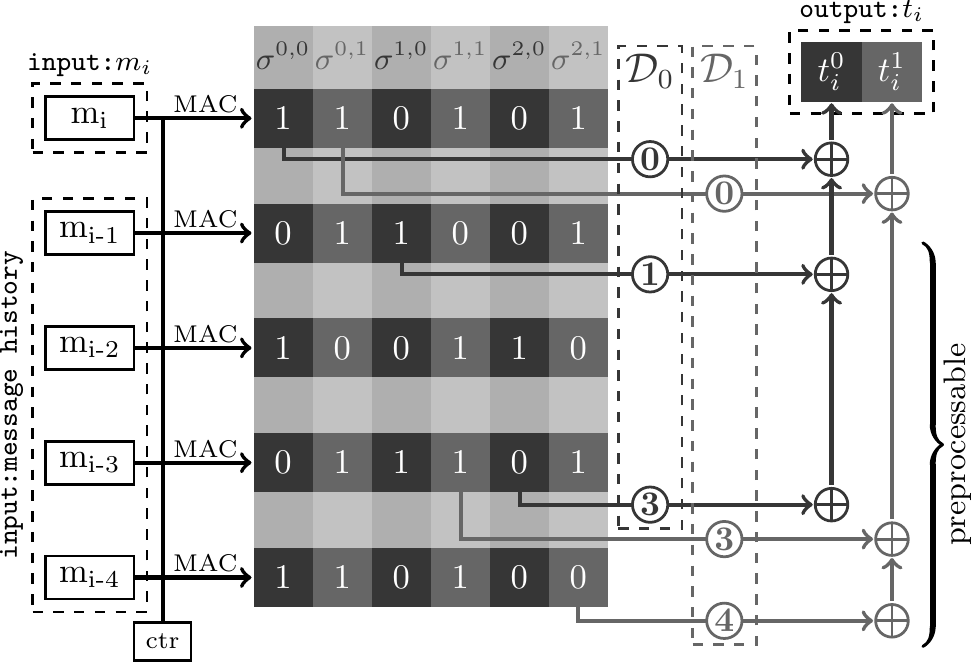}
	\caption{ 
		\name computes traditional MACs~(only 6\,bits shown here) for a message and derives aggregated and compressed tags through efficient XOR operations.
	}
	\label{fig:proposal}
\end{figure}

Orienting ourselves on the good performance results of CuMAC~\cite{2020_li_cumac, 2021_li_cumacs}, \name, in contrast to Whips~\cite{2020_armknecht_promac} and Mini-MAC~\cite{2017_schmandt_minimac}, computes tags using an aggregation procedure for traditional MACs instead of defining how tags are directly derived from the recent message history.
In a nutshell, \name{} thus operates as outlined in Figure\,\ref{fig:proposal}, where exemplarily 6-bit traditional MACs are first computed and then compressed into 2-bit ProMACs tags.

In more detail, to generate $t_i$ for a message $m_i$, \name first computes $\sigma_i$ as a traditional MAC~(\eg \code{HMAC-SHA-256-128}~\cite{2006_bellare_hmac}) over $m_i$ and a counter $ctr$, \ie $\sigma_i = HMAC(m_i, ctr)$.
The counter is initialized with $0$ and incremented after each message to protect against replay attacks.
\name computes each bit of the tag $t_i$ individually since each bit depends on a unique set of past messages~(\cf\ Section\,\ref{sec:design-bit-dependencies}).
For pseudorandomly selecting these bit dependencies $\mathfrak{D}$, a pre-shared secret is used.
Subsequently, \name derives $t_i$ from $\sigma_i$ and from the MACs of messages in the recent past as follows.

In the following, we refer to the $j$-th bit of $t_i$ as $t_i^j$, \ie $t_i=t_i^0||\dots||t_i^{|t_i|-1}$, where $|t_i|$ denotes the bit-length of $t_i$.
\name also splits $\sigma_i$ into its individual bits, by first spitting $\sigma_i$ into $\lceil|\sigma_i|/|t_i|\rceil$ parts.
We then denote $\sigma_i^{a,b}$ as the $b$-th bit of the $a$-th part of $\sigma_i$, \ie $\sigma_i^{a,b}$ is the bit at position $(a\cdot|t_i|+b)$ of $\sigma_i$.
\name then computes each bit $t_i^j$ of the final tag $t_i$ of message $m_i$ using the bit dependencies $\mathcal{D}_j$ for this bit as follows:
\begin{equation*}
	\centering
	t_i^j = \bigoplus_{0\leq n<|\mathcal{D}_j|} \sigma_{i-\mathcal{D}_j[n]}^{n,j}
\end{equation*}
with $\mathcal{D}_j[n]$ representing the $n$-th entry of $\mathcal{D}_j$.
At the start of a message stream, missing values are initialized with $0$.
Consequently, each bit $t_i^j$ of $t_i$ depends exactly on those messages defined in the corresponding bit dependencies $\mathcal{D}_j \in \mathfrak{D}$ and each bit $\sigma_i^{a,b}$ is included in exactly one tag.

To speed up calculations, \name{} partially caches previous messages' tags
until they are fully depleted, \ie all bits have been incorporated into tag computations of subsequent messages.
Furthermore, to reduce latency when computing the tag of a certain message, all but one XOR operation (incorporating the bit dependencies of this message into all tag bits at once) can be preprocessed since this processing only relies on bits from previous messages' tags.
Using efficient XOR operations, which are mostly precomputable in idle time,
\name{} is particularly suitable for resource-constrained environments, which are the prime profiteer of ProMACs.

\subsection{Security Discussion}
\label{sec:construction:security}

The security of \name{} follows from its resilience to key recovery attacks and the unforgeability of tags.

\subsubsection{Resilience to Key Recovery Attacks}

By overhearing the communication, an attacker learns strictly less information from channels that use SP-MAC for integrity protection than channels that rely on \name's underlying MAC scheme, as the tags computed by them are needed to compute the SP-MAC tag.
Hence, a key recovery attack against \name is at least as hard as against the underlying MAC scheme.
Thus, the key used to compute the tags cannot be recovered as long as the underlying MAC scheme does not expose a key recovery attack.
The rationale behind this claim is the following:
Given a stream protected with traditional MACs, an adversary can choose arbitrary bit dependencies~(without needing access to the key) to derive the tags that would have been sent by \name{}, similar to the illustration in Figure\,\ref{fig:proposal}.
Thus, any key recovery attack against \name also attacks the underlying MAC protocol, as an adversary only needs to transform the underlying MAC into \name{}'s representation before launching the 
attack.

\subsubsection{Unforgeability of Integrity Protection}

Security of~(traditional) MACs relies on the unforgeability of tags, \ie attackers can neither directly forge tags nor guess the secret key~(\cf\ Section\,\ref{sec:attack:threat-model}).
When considering ProMACs, the integrity of a single message is secured by multiple tags. 
At the same time, a single tag protects multiple messages.
Other than traditional MACs, ProMACs, therefore, have to define their security based on the~(computational) infeasibility of circumventing the integrity protection of a single message.

We assume that the underlying MAC scheme provides a security level of $n$\,bits using an $n$-bit~tag, \ie the probability of guessing 
a tag is not better than $2^{-n}$.
This assumption is expected to hold for common MAC schemes, \eg \code{HMAC-SHA-256-128}~\cite{2006_bellare_hmac}, and eases discussions on \name{}'s security in face of dropped packets.
To show the security of \name{}, we first look at traditional MACs and think of an $n$-bit~tag as $n$ individual $1$-bit~MACs.
Each of these $1$-bit~MACs provides $1$\,bit of security, \ie the probability that an attacker guesses it correctly is $2^{-1}$. 
A message protected by a traditional MAC is transmitted with its $n$ 1-bit~MACs, and if it is not altered, all $1$-bit~MACs can be verified.
For \name{}, this procedure changes as the $1$-bit~MACs are distributed over multiple packets and aggregated using XORs.

This aggregation of multiple MACs itself does not impact security, as XOR-ing multiple MACs still leads to a secure MAC scheme~\cite{1995_bellare_xormac}.
However, combining multiple MACs introduces dependencies on the successful reception of other messages, as a MAC can only be verified if \emph{all} messages protected by the XOR-ed MACs were received unaltered.
Here, \dname{} ensures that these dependencies are staggered, and thus prevents the sandwich attack introduced earlier~(\cf\ Section\,\ref{sec:attack}).
Consequently, a significant number of messages have to be dropped to render a large number of MAC fragments covering a single message unverifiable. 
To illustrate this issue, in case R2-D2 is parameterized for a maximum security loss of 16\,bits per dropped message for a maximum security level of 128 bit and 4~targeted messages could be dropped, \name still achieves a security level of at least 64 bit for the targeted message.
However, \name's selected bit dependencies $\mathfrak{D}$ are derived from a pre-shared secret, thus hiding the strategy to achieve this worst-case attack from third parties.
Consequently, an adversary needs to drop a suspiciously high number of transmissions~\cite{2020_armknecht_promac} to even attempt to circumvent the integrity protection of a single message.

By providing resilience to key recovery attacks and ensuring the unforgeability of integrity protection, \name{} is able to realize secure progressive message authentication.
Most notably, \name{} is the first ProMAC scheme that offers protection against network-level attacks, while still quickly achieving full integrity protection.

\subsection{Performance Evaluation}
\label{sec:construction:eval}

ProMACs are specifically designed for resource-constrained environments, especially for wireless scenarios with high-frequency communication~(\cf~Section\,\ref{sec:motivating-example}). 
In this context, energy consumption is generally a key metric for battery-powered devices.
However, since the bulk of energy of those devices is consumed by the transmission and reception of wireless communication~\cite{radio_power}, the difference between ProMACs schemes is negligible. 
Nevertheless, for this reason, the overall energy cost of ProMACs is significantly lower compared to traditional MACs as they require longer tags and thus have a higher transmission overhead.
Consequently, our evaluation of \name{} focuses on the two most important aspects of these environments:
the computational overhead of tag generation and the memory overhead necessary to track message histories.

\begin{figure}
	\centering	
	\includegraphics[width=\columnwidth]{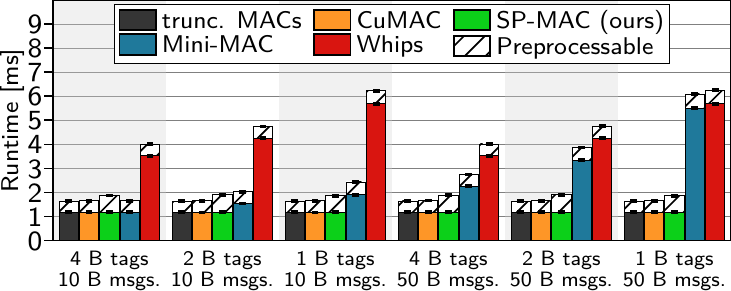}
	\caption{
		\name's tag computation adds only marginal (preprocessable) overhead compared to traditional MACs while twarting network-level attacks against ProMACs.}
	\label{fig:latency-overhead}
\end{figure}

\subsubsection{Computational Overhead of Tag Generation}
\label{sec:construction:eval:computation}

To evaluate the computational overhead of tag generation in \name{}, we implemented a prototype in C for the Contiki-NG~\cite{2004_dunkels_contiki} platform, which is widely used for low-power embedded systems in resource-constrained scenarios~\cite{2017_katsikeas_lightweight, 2018_hiller_antedated, 2019_hiller_tor4iot, 2019_randhawa_energy}.
Our prototype relies on the \code{HMAC-SHA-256-128} implementation of tinyDTLS as an underlying MAC scheme and uses the same \dname{} parametrization as in Section\,\ref{sec:design-bit-dependencies}~(maximal security loss of 32\,bits per dropped packet and 16-bit immediate protection for 4-byte tags).

To compare the performance of \name{} with state-of-the-art ProMAC schemes, we additionally re-implemented\footnote{Re-implementation was necessary as no source code was available.} Whips~\cite{2020_armknecht_promac}, CuMAC~\cite{2020_li_cumac, 2021_li_cumacs}, and Mini-MAC~\cite{2017_schmandt_minimac} for the same platform and underlying MAC scheme.
Furthermore, we use a \code{HMAC-SHA-256-128} MAC as baseline reference~(we truncate the MAC for a fair comparison, although the full MAC needs to be computed anyway).
To encourage further research, we will make all of our ProMAC implementations for Contiki-NG available to the research community. 

With our implementations, we measure the time required to generate one tag of length 1, 2, and \unit[4]{byte}, respectively, for 10 and \unit[50]{byte} long messages on a Zolertia RE-Mote embedded device~(ARM Cortex-M3\,@\,32MHz, 16\,kB RAM).
All MAC schemes are parameterized for 128-bit security, except for the truncated MAC~(baseline).
We performed each measurement 30 times and report on the mean over these runs with 99\% confidence intervals.

The measurements presented in Figure\,\ref{fig:latency-overhead} establish a baseline of 1.17\,ms for computing a traditional~(truncated) MAC, irrespective of tag size and message length
\footnote{This baseline can further be improved if deemed necessary, \eg with hardware acceleration. Resulting performance savings carry over to all four ProMAC schemes.}.
In contrast, the runtimes of Whips and Mini-MAC depend on the size of their internal state.
For Mini-MAC, this state increases with growing message sizes and shrinking tags, ranging from 1.18\,ms~(4-byte long tag and 10-byte long message) to 5.52\,ms~(1-byte long tag and 50-byte long message).
While Whips' processing overhead is independent of message sizes, it also increases for shrinking tags from 3.52\,ms for 4-byte long tags to 5.69\,ms for 1-byte long tags.
Whips starts with a higher processing overhead, as it calls the underlying MAC function twice, once to compute a new substate and once to derive the actual tag.

On top of the baseline for truncated MACs, CuMAC introduces one additional non-preprocessable XOR operation, resulting in nearly identical runtime as truncated MACs.
\name has the same online performance but adds a marginal 0.23\,ms preprocessing overhead to realize R2-D2~(\cf\ Section\,\ref{sec:construction:design}) and protect against network-level attacks. 
As expected, neither \name's nor CuMAC's runtime is noticeably influenced by tag sizes or message length.

These results show that \name{} not only efficiently protects against network-level attacks but even operates at least as resource-conscious as existing ProMACs.
Notably, \name{}'s performance closely aligns with traditional MACs, showing that the benefits of ProMACs~(\cf\ Section\,\ref{sec:applications}) can be realized without increased latency and with only a minor increase in consumed processing power.

\subsubsection{Memory Overhead for Keeping Integrity State}

\begin{figure}
	\centering	
	\includegraphics[width=\columnwidth]{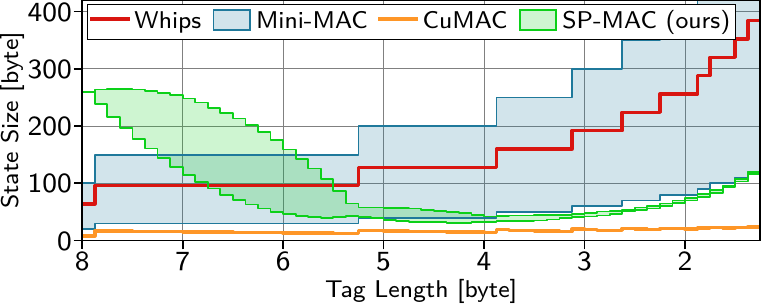}
	\caption{ 
		Despite tracking longer histories to thwart attacks, \name's memory footprint is in the same order as vulnerable ProMAC schemes.
		Note the decreasing x-axis that represents increasing space savings by ProMACs.
	}
	\label{fig:memory-overhead}
\end{figure}

In contrast to traditional~(truncated) MACs, ProMACs inherently have to keep state on past messages for the computation of future tags~(\cf\ Section\,\ref{sec:formal-definition}).
To evaluate the impact of this state keeping on the memory consumption of resource-constrained devices, we evaluate the size of \name's state in relation to the tag lengths and, again, compare it to Whips, CuMAC, and Mini-Mac.
To avoid potential bias due to implementation decisions, we conduct a theoretical memory analysis that is independent of our re-implementations of current ProMAC schemes.
We parametrize all schemes to provide at least 128-bit security, with security loss limited by tag sizes, which is the worst case for SP-MAC \wrt memory overhead.

The results of our analysis
are visualized in Figure\,\ref{fig:memory-overhead}.
First, the memory overhead of Whips depends on the number of tags required for the desired security level~(security level divided by tag length).
Its overhead ranges from \unit[64]{bytes} for 8-byte long tags 
to \unit[416]{byte} for 10-bit long tags.
As the overhead of Mini-MAC additionally depends on the message length, we study the overhead for messages with lengths between 10 and \unit[50]{byte} and find that the memory overhead of Mini-MAC follows the same trend as Whips. 
However, Mini-MAC's overhead depends on the message length, such that the resulting overhead spans an area around Whips' overhead.
Exemplary, for a tag length of \unit[4]{byte}, the overhead of Mini-MAC ranges from \unit[40]{byte} to \unit[200]{byte}.
In contrast, CuMAC has a low and relatively constant memory overhead since the state stored per message decreases proportionally to the growing number of messages that are aggregated within one tag as tags vary in length.

Similar to CuMAC and Whips, the memory overhead of \name{} does not depend on the message length.
However, it is influenced by the \emph{random} selection of bit dependencies~(\cf\ Section\,\ref{sec:design-bit-dependencies}), as well as the interfering effects of tag length reductions~(fewer bits per tag have to be precomputed) and the exponentially growing size of Golomb Rulers of increasing orders~(tag precomputation has to start earlier).
We thus display the range between the best and worst possible bit dependencies \wrt~their memory overhead.
As shown in Figure\,\ref{fig:memory-overhead}, this range is negligible for small tags and increasing for larger tags because the number of short Golomb Rulers with two elements~(used in tags longer than 42 bit) is limited.

Overall, the memory footprint of \name{} is well-manageable regardless of tag lengths, even for devices with scarce resources~(\cf\ Section\,\ref{sec:construction:eval:computation}).
Moreover, even in the worst-case scenario for \name{} in terms of memory overhead, its memory footprint is in the same order as current ProMAC schemes, all of which are vulnerable to the presented sandwich attack~(\cf~Section~\ref{sec:attack}).

\section{Related Work}

This paper identifies an inherent vulnerability of existing ProMAC schemes and proposes an efficient, flexible, and secure way to achieve strong integrity protection in resource-constrained environments. 
Besides the directly related prior work on progressive message authentication (covered in Section\,\ref{sec:background}), different streams of research tackle this challenge.
Competing technologies to reduce authentication tag sizes are truncated MACs~\cite{1995_preneel_mdx, 2017_wang_survey, 2011_schweppe_car2x, 2008_szilagy_flexible}, stateful MACs~\cite{2020_armknecht_promac}, or aggregated MACs~\cite{1995_bellare_xormac, 1997_gennaro_sign, 2008_katz_aggregated-mac, 2010_eikemeier_history, 2012_kolesnikov_mac}, which, however, only achieve reduced security or cannot cope with lossy channels. 
Note that our approach can also be used \emph{conservatively} as an alternative to aggregated MACs as discussed in the Appendix\,\ref{app:aggregated-scheme}.

Beyond ProMACs, several other message authentication schemes using optimistic security have been proposed in similar resource-constrained environments as those considered in this paper. 
Only reacting to multiple invalid short tags within a short time span~\cite{2008_szilagy_flexible, 2010_szilagyi_low,2012_szilagyi_phd} does, however, not protect against the manipulation of a few selected messages.
The optimistic and selective use of aggregated MACs~\cite{2017_castellanos_retrofitting2} does require reliable transmissions and leads to high authentication delays, two limitations that ProMACs address.
Relying on external devices to verify broadcasted information~\cite{2008_nilsson_efficient} is only applicable with costly asymmetric cryptography that leads to unacceptable delays and bandwidth overhead in many scenarios.

Targeting to avoid all communication overhead for authentication, different approaches propose sender identification based on unique physical characteristics of the transmission signal~\cite{2018_kneib_scission, 2020_kneib_easi,2014_murvay_source,2018_choi_identifying} or host behavior~\cite{formby_2020, ahmed_2021}.
In contrast to the reliable and deterministic nature of MACs, these approaches are, however, restricted to single-hop transmissions 
and often produce a significant number of false positives.
Similarly, approaches to transparently retrofit integrity protection into legacy protocols hardly elongate packets, but therefore result in reduced security (\eg by using truncated MACs)~\cite{2004_wright_retrofitting1, 2011_herrewege_retrofitting3, 2011_schweppe_car2x, 2008_szilagy_flexible, 2016_radu_leia}.
ProMACs can replace truncated MACs in these protocols and thus improve security guarantees.

Specifically focusing on \emph{multicast} message authentication in constrained environments, related work addresses the cost of asymmetric cryptography by splitting integrity verification over multiple packets~\cite{2001_golle_authenticating} or by adapting symmetric cryptography to the multicast setting~\cite{2000_perrig_multicast-mac, 2001_perrig_efficient, 2002_perrig_spins, 2004_liu_utesla, 2005_challal_rlh, 2007_wu_fast}.
For example, TESLA~\cite{2000_perrig_multicast-mac, 2001_perrig_efficient, 2004_liu_utesla} achieves this through time-delayed key disclosure.
BECAN~\cite{2012_rongxing_becan}, in contrast, improves bandwidth consumption for cooperative authentication scenarios where multiple devices must authenticate a single message.
ProMACs address deficits in unicast message authentication, but their loss-tolerance shows potential to also improve multicast authentication based on symmetric cryptography.

To improve the performance of symmetric cryptography on resource-constrained devices, different approaches propose to leverage special lightweight ciphers~\cite{2014_gong_tulp, 2014_mouha_chaskey, 2016_luykx_mac}, use hardware acceleration~\cite{2018_hiller_antedated,2017_katsikeas_lightweight,2017_yang_hardware}, or preprocess cryptographic operations~\cite{2018_hiller_antedated, bpmac, mergemac}.
As our work is agnostic to the underlying MAC scheme, these performance improvements conceptually carry over to \name{}.

\section{Conclusion}

Progressive message authentication codes~(ProMACs) promise the compression of authentication tags while preserving the strong security of traditional MACs by partly offloading integrity protection into the near future.
Contrary to prior beliefs, we show that ProMACs cannot cope particularly well with lossy channels, preventing their deployment in many wireless scenarios:
The generation and verification of tags depend on a sliding window of past messages, providing the foundation for our \emph{sandwich attack}, in which the integrity protection of a whole message sequence is rendered void if merely two packets are dropped.
Therefore, we consider it imperative to rethink how transmission failures influence the integrity protection of neighboring messages.
With this in mind, we propose randomized and resilient dependency distributions~(\dname), which takes advantage of
\begin{enumerate*}[label=(\roman*)]
	\item optimal message dependencies,
	\item parameterized security guarantees,
	\item randomized bit dependencies, and
	\item optional immediate protection bits.
\end{enumerate*}
Our evaluation shows that \dname{} significantly increases the resilience of ProMACs to lossy channels to unleash their full potential.
At the same time, \dname achieves full integrity protection with comparable delays to current ProMAC schemes.
To take advantage of \dname and realize a secure and resource-conscious ProMAC scheme, we propose \name{}
that builds upon the proven security provided by traditional MACs, aggregating and distributing those across multiple messages using efficient XOR operations.
\name is thus not only resilient to lossy channels and sophisticated network-level attacks but also operates as resource-conscious as state-of-the-art ProMAC schemes.

\begin{acks}
	Funded by the Deutsche Forschungsgemeinschaft (DFG, German Research Foundation) under Germany's Excellence Strategy -- EXC-2023 Internet of Production -- 390621612.
	We thank Misha Lavrov for pointing us to Golomb Rulers as well as the anonymous reviewers and our shepherd Mridula Singh for their fruitful comments.
\end{acks}

\bibliographystyle{ACM-Reference-Format}
\bibliography{paper}

\appendix

\section{Appendix}

\subsection{Implications of the Sandwich Attack}
\label{app:attack-in-practice}

As shown in this paper, current ProMAC schemes suffer from a common vulnerability inherent to their design:
Since integrity protection is distributed among \emph{consecutive} messages, the (forced) loss
of two messages that are less than the  
message history apart already disables integrity protection for all messages in between.
Thus, with minimal and unsuspicious interference, an attacker can covertly remove authenticity from multiple ProMAC-protected messages.
Depending on how ProMACs are deployed, this vulnerability leads to effective denial-of-service attacks or even false data injections.

\subsubsection{Reverting All Suspicious Traffic} 
ProMACs promise to enable the optimistic processing of messages upon reception based on  reduced initial security.
In the unlikely event of a retroactive detection that a message could not be authenticated, the effects of a potentially malicious message have to be reverted. %
If ProMACs operate on a high-reliability link with hardly any packet loss, this mode of operation is reasonable.
However, in the presence of a jamming attack or a less reliable channel, current ProMACs schemes cannot ensure a stable operation due to frequent rollbacks.
Consider a \emph{selective jammer} that listens to the communication channel to identify ``interesting'' packets, \ie those that need to be suppressed to launch a sandwich attack, and then deliberately distorts these packets~\cite{2011_wilhelm_reactive-jamming, 2016_hamza_jamming-survey, 2017_aras_jamming}. 
To implement such a jammer, an attacker has to control a device that
\begin{enumerate*}[label=(\roman*)] 
	\item is in range of the targeted communication and
	\item can actively jam ongoing communication.
\end{enumerate*}

\begin{figure}
	\centering	
	\includegraphics[width=\columnwidth]{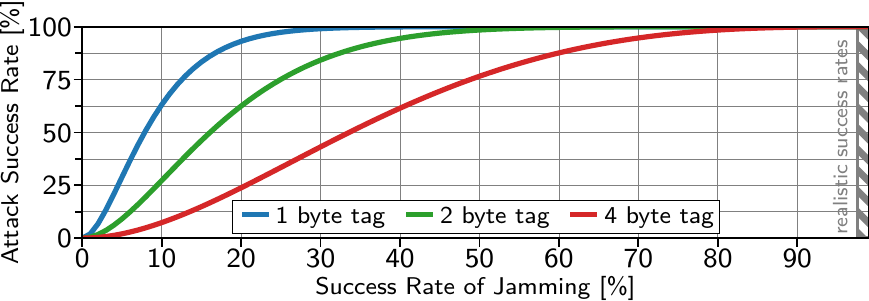}
	\caption{%
		How well a selective jammer can execute the sandwich attack depends on the effectiveness of jamming.
		In realistic scenarios, the attack can be launched reliably even from an imperfect jammer.
	}
	\label{fig:active}
\end{figure}

In Figure\,\ref{fig:active}, we study how jamming capabilities translate to success rates for the sandwich attack.
A successful attack means that a considered packet could not be authenticated, which in this %
deployment scenario forces the entire system to roll back to %
the point at which the targeted message has been optimistically processed.
To this end, we assume that an attacker attempts to make exactly one message unauthenticable by jamming the neighboring packets influencing this message's integrity. 
We consider three tag sizes~(1, 2, and 4\,bytes) and analytically compute the attack success rate for varying likelihoods of successful jamming.
Our results show that even an imperfect jammer can execute the sandwich attack reliably.  
Furthermore, smaller tags are more susceptible to selective jammers (for identical security levels), as their larger sliding windows~(which are inversely proportional to tag sizes) give an attacker more opportunities to jam packets.

To put our results into perspective, we highlight~(\,\raisebox{.7mm}{\fcolorbox{gray}{gray}{\rule{0pt}{2pt}\rule{2pt}{0pt}}}~in Figure\,\ref{fig:active}) the practical likelihood of successful selective jamming~\cite{2011_wilhelm_reactive-jamming,2017_aras_jamming}.
For IEEE 802.15.4, used in common wireless protocol stacks for constrained devices, selective jamming has proven effective in covertly dropping packets with success rates between 97.6\% and 99.9\%~\cite{2011_wilhelm_reactive-jamming}.
Similar numbers have been reported for LoRaWAN, used for energy-efficient long-range IoT communication, where selective jamming using commodity hardware shows success rates between 98.7\% and 99.9\%~\cite{2017_aras_jamming}.
These success rates can further be improved if transmission times can be predicted, \eg when TDMA is used on the medium access layer~\cite{2016_hamza_jamming-survey}.
Considering these numbers, our analysis shows that active jammers can execute the sandwich attack with success rates above 99.9\% in realistic settings.
Consequently, in this deployment, a jammer can cripple a ProMAC-protected communication link with selective, and thus stealthy, interference.

Furthermore, current ProMACs cannot be exposed to harsh environments, such as 
 \eg \ac{ICS}~(\cf~Section~\ref{sec:motivating-example}) with realistic error rates of 1 to 10\,\%~\cite{2008_thonet_zigbee,2013_wei_rtwifi}, without restrictions.
To validate this claim, we simulate two wireless communication channels~(``low-error'' and ``high-error'') based on the Gilbert-Elliot~(G-E) model, commonly used to simulate wireless channels based on a Markov chain with two states~\cite{2019_daailva_mac-model-survey}.
In the G-E model, the two states are used to encode a ``good'' and ``bad'' channel state, with corresponding packet drop probabilities $\code{err}_\code{good}$ and $\code{err}_\code{bad}$.
The parameters \code{p} and \code{r} define the probability for switching from ``good'' to ``bad'' and vice versa. %
We parameterize the G-E models as summarized in Table\,\ref{tab:model} based on recommendations from the literature~\cite{2021_haenel_ge-params}, to represent packet error rates of approximately $1\%$~(low-error) and $10\%$~(high-error), which are realistic for scenarios envisioned for ProMACs~\cite{2008_thonet_zigbee,2013_wei_rtwifi}.

\begin{figure}
	\centering	
	\includegraphics[width=\columnwidth]{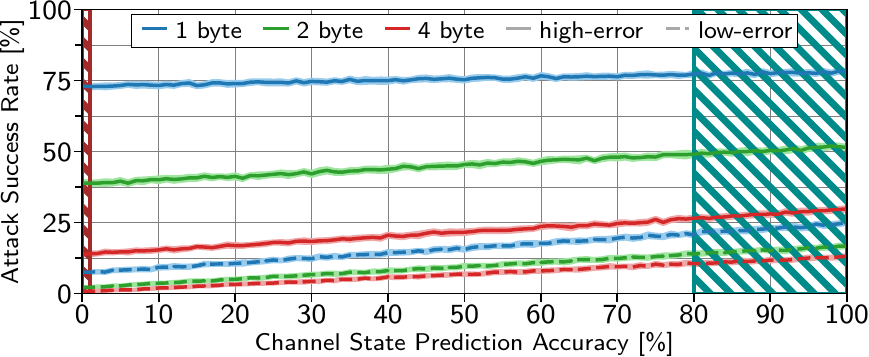}
	\caption{ 
		Even attackers without jamming capabilities can execute the sandwich attack, especially if they can predict the state of the communication channel well enough.}
	\label{fig:passive}
\end{figure}

Using these two channel models, we first investigate the number of unauthenticatable packets for varying tag sizes~(1, 2, and 4\,bytes).
Figure\,\ref{fig:passive} reports on the mean attack success rate over 30~Monte Carlo simulations covering 1000~attacks with 99\%~confidence intervals.
A successful ``attack'' again means that the considered packet could not be authenticated.
We found that the overall channel quality, as well as the used ProMAC tag length, influence the attack's success rate.
In particular, we observe~(\,\raisebox{.7mm}{\fcolorbox{brown}{brown}{\rule{0pt}{2pt}\rule{2pt}{0pt}}}~in Figure\,\ref{fig:passive}) that between 0.6\% or 73.1\% of messages do not have any verifiable integrity protection, depending on the overall channel quality and tag length.
These results indicate the number of overall unauthenticable packets for current ProMAC schemes over lossy channels.

\begin{table}
	\centering
	\small
	\begin{tabularx}{\columnwidth}{l Y Y Y Y Y}
		& \multicolumn{4}{c}{Model Parameters~(\%)} & resulting \\
		\addlinespace[-0.25em]
		\cmidrule(lr){2-5}
		\addlinespace[-0.3em]
		Channel &  \code{p} &  \code{r} & $\code{err}_\code{good}$ &   $\code{err}_\code{bad}$ & avg. PER\\
		\toprule
		\textbf{low-error} &  0.5 &   75.9 &   1.1 &   62.4 &   1.50\% \\
		\textbf{high-error} &   3.2 &   83 &   6.8 &   78.8 &   9.47\% \\ %
		\bottomrule
	\end{tabularx}
	
	\caption{ Our parametrization of the Gilbert-Elliot models of a realistic low- and high-error link based on recent longitudinal measurements of industrial networks~\cite{2021_haenel_ge-params}.}
	\label{tab:model}
\end{table}

\balance
\subsubsection{Reacting Only to Explicitly Detected Attacks} 

Previous publications on ProMACs~\cite{2020_armknecht_promac, 2020_li_cumac, 2021_li_cumacs, 2017_schmandt_minimac} imply a second deployment scenario, where rollbacks only take place after a manipulation is \emph{explicitly} detected.
However, in this scenario, an attacker can inject malicious traffic into a data stream that is not detected as such by jamming the neighboring packets with a high success rate~(\cf~Figure~\ref{fig:active}).
Alternatively, a less powerful attacker can abuse the naturally occurring unauthenticatable packets on a link with higher error rates.
If attackers are even able to predict a bad channel state, they can increase the chances of their data injection not being detected.
To illustrate this behavior, Figure~\ref{fig:passive} increase the channel state prediction accuracies (specifying the likelihood of the channel being in the ``bad'' state) for the attacker on the x-axis.
After predicting a bad channel, the attacker waits for one additional transmission before injecting a forged message.
The attack is successful if the two windows, starting and ending with the forged message, each contain at least one transmission failure.

In practice, such prediction accuracies can be upwards of 80\% if the attacker is in the vicinity of the receiver~($<$\,\unit[1]{m} apart), as indicated by practical measurements~\cite{2010_zhu_correlation}.
Assuming an accuracy of 80\%, an attacker can successfully launch a sandwich attack in 10.4\% to 75.4\% of attempts, depending on the link quality and tag lengths~(\,\raisebox{.7mm}{\fcolorbox{darkcyan}{darkcyan}{\rule{0pt}{2pt}\rule{2pt}{0pt}}}~in Figure\,\ref{fig:passive}). 
Overall, the prospects of ProMACs are extremely desirable~(\cf\ Section\,\ref{sec:applications}), but even a moderate error rate on the transmission medium leads to significant risk of false data injections or crippled communication channels.
Meanwhile, smaller tags with larger sliding windows favor the sandwich attack exactly for those scenarios that benefit the most from ProMACs.

\subsection{SP-MAC as an Aggregated MAC Scheme} 
\label{app:aggregated-scheme}

While this paper focuses on ProMACs that process messages \emph{optimistically} upon reception, ProMACs can also be used \emph{conservatively} as an alternative to aggregated MACs.
In such a scenario, messages would only be processed once they are fully authenticated and discarded otherwise.
This scenario, however, does not align with the low-latency properties attributed to ProMACs and has thus not been considered in the literature so far.
Still, we demonstrate how our later improvements to ProMACs allow them to outperform aggregated MAC schemes in such deployment.

\begin{figure}
	\centering	
	\includegraphics[width=\columnwidth]{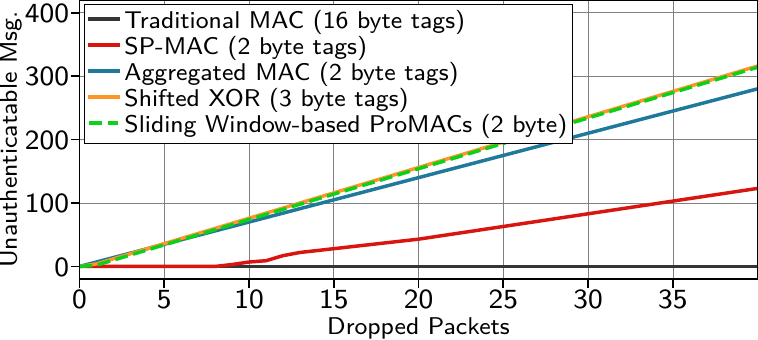}
	\caption{
		While traditional aggregated MAC schemes are immediately disrupted by dropped packets, \name, as aggregated MAC schemes, can tolerate some dropped packets and reduce the effects of ongoing DoS attacks in comparison to traditional schemes.
	}
	\label{fig:dos}
\end{figure}

In an aggregate MAC scheme, the authentication of a batch of messages is aggregated into a single tag that is sent to the receiver after the last message of the batch has been transmitted.
Those schemes only process messages once a certain bit level of security is achieved and thus do not enable the same low latency processing as ProMACs.
Still, many scenarios do not require low latency, and current aggregated MAC schemes~\cite{1995_bellare_xormac,1997_gennaro_sign,2008_katz_aggregated-mac,2010_eikemeier_history,2012_kolesnikov_mac} are a well-proven method to reduce the overhead of traditional message authentication in such cases.
However, similar to ProMACs, aggregated MAC schemes are susceptible to network-level interference: Attacks similar to the one presented against ProMAC schemes~(\cf~Section~\ref{sec:attack}) allow for an attacker to drop a few select packets having a cascading effect on the verifiability of neighboring packets~\cite{2012_kolesnikov_mac}, which finally results in many discarded transmissions.
In the following, we show how \name, our proposed ProMAC scheme, is also resilient to network-level interference when used as an aggregated MAC scheme.
For our analysis, we compare different aggregated MAC schemes based on how many messages are discarded due to the lack of integrity protection in the presence of an attacker that drops selected packets.
As a baseline, we use traditional MACs, where the receiver can verify the integrity of each received packet.
As aggregated MAC scheme, we first use a traditional scheme~\cite{2008_katz_aggregated-mac,2010_eikemeier_history}.
Additionally, we consider schemes that are aggregated based on shifted XORs~\cite{2012_kolesnikov_mac}, an aggregation mechanism that protects, at the cost of slightly longer tags, against DoS attacks where the attacker is only able to selectively drop certain packets.
In a simulation, we compare these schemes against sliding window-based ProMAC schemes~(Whips~\cite{2020_armknecht_promac}, CuMAC~\cite{2020_li_cumac,2021_li_cumacs}, and Mini-MAC~\cite{2017_schmandt_minimac}) and our proposed ProMAC scheme (\name).
We consider short authentication tags~(2 bytes in most cases) and parameterize \name to a maximum security loss of 16\,bit per dropped packet.
For the five considered approaches, we let an attacker drop an increasing number of selected packets chosen to maximize the number of packets discarded by the receiver.
We consider a message authentic, \ie it does not have to be discarded by the receiver if it reaches at least a security level of 32\,bit.

The results of our analysis are shown in Figure~\ref{fig:dos}.
For traditional MACs, we see, as expected, that no message loses its integrity protection while being received.
Aggregated MACs, on the other hand, aggravate the effect of an attack because a single dropped packet invalidates all messages authenticated in the corresponding batch~(8 in this case).
Even the Shifted XOR aggregation scheme does not improve resiliency, as it was specifically designed to protect against attacks that cannot drop all packets.
In fact, the Shifted XOR aggregation scheme performs even worse than aggregated MACs for an attacker that can drop arbitrary packets.
Sliding window-based ProMACs exhibit similarly poor results, as an attack can invalidate a sequence of messages by dropping the two packets at the edge of this sequence~(\cf~Section~\ref{sec:attack}).

Regarding \name, we observe a completely different behavior than that of other aggregated MAC schemes.
First, the initial packet drops do not invalidate the integrity protection of any messages transmitted in other packets.
The security properties of \dname, on which \name is built, ensure that only a certain fraction of each message's integrity protection (16\,bit in our parameterization) depends on the successful reception of any other packet.
Furthermore, \name's optimized dependencies on surrounding packets also provide protection against ongoing attacks.
While \name cannot ensure that all messages' integrity protection can be verified, it still cushions the effects of the DoS attack by limiting the number of unauthenticatable messages to less than half of what aggregated MACs achieve.
Thus, while exact numbers still depend on \name's specific parameterization (\eg targeted security level or maximum security loss), our analysis shows that \name significantly outperforms competing aggregated MAC schemes in scenarios with packet loss caused by either lossy channels or an active attacker.
Consequently, \name is not only the first ProMAC scheme that is resilient to sandwich attacks but also improves the state-of-the-art of aggregated MAC schemes.

\begin{figure*}
	
	\begin{subfigure}{\columnwidth}
		\centering
		\includegraphics[width=\textwidth]{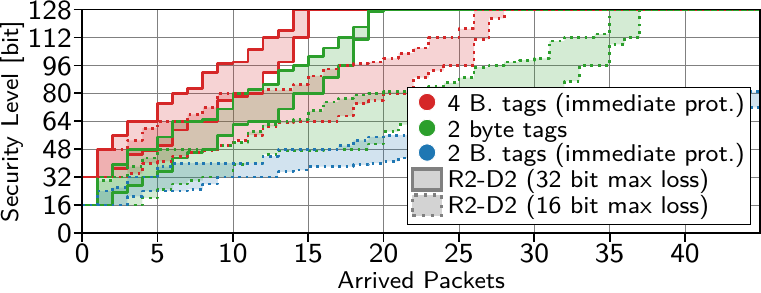}
		\caption{ Higher resilience to network-level attackers inherently increase the delay until full authenticity can be reached, but keeps them within a range that is still manageable in many scenarios. }
		\label{fig:add_tags}
	\end{subfigure}
	\hfill
	\begin{subfigure}{\columnwidth}
		\centering
		\includegraphics[width=\textwidth]{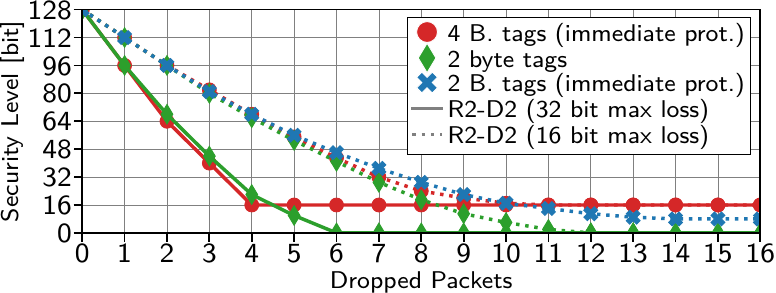}
		\caption{
			By reducing the maximally lost bit security per dropped packet, the resilience to network-level attacks increases accordingly as has been proven for \dname's dependency selection.
		}
		\label{fig:add_drop}
	\end{subfigure}
	
	\caption{ \dname's flexible parameterization enables a fine-granular and controllable trade-off between security, authentication delay, and message overhead.}
	\label{fig:add}
\end{figure*}

\subsection{Proofs of Security Properties}
\vspace*{1mm}
\subsubsection{Security Properties of Golomb Rulers-based dependencies}~
\label{app:proof-gr}

\vspace*{2mm}
\textit{\textsc{Proposition\,1.}}
By dropping a message $m$ of a ProMAC-protected stream, any other message's integrity protection is reduced by at most the security provided by one tag iff the dependency $\mathcal{D}$ is a Golomb Ruler.\\[-.5em]

\textit{ \textsc{Proof.}} 
Without loss of generality, we investigate the authenticity of the message $m_i$.
The integrity of $m_i$ is protected by tags $ \{ t_{i+d}| d\in \mathcal{D} \}  = \mathcal{P}$.
We prove our claim by contradiction. 
Therefore, we assume that there exists $m_j (j\neq i)$, such that at least two tags from $\mathcal{P}$ become unverifiable if $m_j$ is dropped.
These two tags have the form $t_{i+d}\in \mathcal{P}$ and $t_{i+d'}\in \mathcal{P}$ $(d\neq d')$.
If both of these tags become unverifiable because $m_j$ is dropped, $ m_j $ has to be included in the intersection $\mathcal{I} = \{ m_{i+d-\delta}| \delta \in \mathcal{D} \} \cap \{ m_{i+d'-\delta'}| \delta' \in \mathcal{D} \}$ of the messages that are required to compute these tags.
Since we assumed that $m_j\in\mathcal{I}$, this requires the existence of $d,d',\delta, \delta' \in \mathcal{D}$, such that $d-\delta=d'-\delta'$.
However, exactly when $\mathcal{D}$ is a Golomb Ruler, this only holds if both differences equal $0$, \ie $d = \delta$ and $d' = \delta'$.
However, if $d-\delta=d'-\delta'=0$, then it has to hold that $\mathcal{I} = \{ m_i \}$, which means that $m_j (i\neq j) \notin \mathcal{I}$.
Thus, there cannot exists any message $m_j$ that, if dropped, invalidates multiple tags authenticating $m_i$. $\square$\\[-.9em]

\vspace*{1mm}
\subsubsection{ Security Properties of $g$-Sidon Set-based dependencies}~
\label{app:proof-ss}

\vspace*{2mm}
\textit{\textsc{Proposition\,2.}}
Using $g$-Sidon Sets as ProMAC dependencies $\mathcal{D}$ guarantees that any message's security level is reduced by at most the integrity protection provided by $g$ tags for any dropped message.\\[-.5em]

\textit{\textsc{Proof.}}
Without loss of generality, we investigate the authenticity of the message $m_i$.
The integrity of $m_i$ is protected by tags $ \{ t_{i+d}| d\in \mathcal{D} \}  = \mathcal{P}$.
We prove Proposition 2 by contradiction. 
Therefore, we assume that there exists $m_j (j\neq i)$, such that at least $g+1$ distinct tags from $\mathcal{P}$ become unverifiable if $m_j$ is dropped.
These $g+1$ tags have the form $t_{i+d}\in \mathcal{P}$, with a distinct $d$ for each tag.
If all of these $g+1$ tags become unverifiable because $m_j$ is dropped, $ m_j $ has to be included in the intersection $\mathcal{I} = \{ m_{i+d-\delta}| \delta \in \mathcal{D} \} \cap \{ m_{i+d'-\delta'}| \delta' \in \mathcal{D} \}$ of the messages that are required to compute these tags.
Since we assumed that $m_j\in\mathcal{I}$, this requires the existence of $g+1$ distinct $\delta$-$\delta'$ pairs, such that $d-\delta=d'-\delta'$, with $d,d',\delta, \delta' \in \mathcal{D}$.
However, exactly when $\mathcal{D}$ is a $g$-Sidon Set, there exist by definition at most $g$ distinct $\delta$-$\delta'$ pairs~(\cf~Section~\ref{sec:param-sec-guarantees}).
Thus, there cannot exist a message $m_j$ that, if dropped, invalidates more than $g$ tags that authenticate $m_i$. $\square$

\subsection{Looking at Additional R2-D2 Parameters }
\label{app:more-results}
 
In this paper, we mainly considered a security loss of 32~bits per dropped packet as an acceptable security guarantee.
Indeed, in Section~\ref{sec:security-properties}, we saw how exactly such a parameterization provides adequate security in a range of practical examples while keeping the delay to achieve full authentication within an acceptable range.
However, we also observe that these guarantees might not provide sufficient resilience to network-level attacks.
Meanwhile, \dname offers flexible parameterization, which enables finding an application-specific balance between security, authentication delay, and tag size.
To further illustrate how these trade-offs for different parameters, we show three additional parameters sets that reduce the security loss per dropped packet to at most 16~bits.

Since 1-byte tags already lead to significant delays if security loss was parameterized to 32~bit~(cf.~Figure\,\ref{fig:r2d2_tags}), we consider 4 and 2~byte long tags.
For those tags, we reserve half of them, \ie 16 and 8\,bit respectively, for immediate protection bits, as those ensure that no network-layer attacker can remove all integrity protection for any single message.
Additionally, we include a 2-byte tag that does not have any immediate security bits.
Figure\,\ref{fig:add} reports on the results for these new parameters (dotted lines) as well as the \dname parameters previously evaluated in Section~\ref{sec:design-bit-dependencies} for the same tag lengths.
In Figure\,\ref{fig:add_tags}, we observe that delays to achieve full authentication are prolonged by reducing maximally accepted bit security loss per dropped packet.
While the delays for tags with 16~progressive security bits are at most 37~messages, which is acceptable for many applications~(\cf~Section~\ref{sec:security-properties}), 
delays for 2-byte tags with 8~bits of immediate security reach up to 75~messages.
Thus, such short tags, with the additional benefit of never losing all authenticity due to a network-level attacker, introduce considerable delays.
If these delays are not acceptable and ProMACs are still desirable, some other trade-offs must be made in such situations (\ie slightly longer tags or less resilience against network-layer attackers), since \dname's delays are provably optimal and can not be further improved without such trade-offs.

Finally, Figure\,\ref{fig:add_drop} shows how the increased resilience to dropped packets manifests itself in practice.
As expected, at the cost of longer authentication delays, \dname can significantly improve ProMACs' resilience to packet loss.
Additionally, this figure illustrates what has been previously proved for \dname under its various parameters, \ie the achieved bit security never reduces by more than 16 bits per dropped packet for the newly evaluated parameters~(dotted lines).
All in all, we can conclude that the provably optimal dependency distributions of \dname for its core parameters (maximum bit security per dropped packet, tag length, and immediate security bits) are thus flexibly adaptable to the needs of many realistic scenarios to realize ProMACs with adequate resilience to network-level attackers.

\end{document}

%% file: include.tex
\usepackage{xcolor}
\usepackage[detect-weight=true,detect-family=true]{siunitx}
\usepackage{xspace}
\usepackage{tikz}

\usepackage{soul}

\usepackage{booktabs}
\usepackage{multirow}

\usepackage{units}

\definecolor{darkgrey}{RGB}{80,80,80}
\definecolor{lightgrey}{RGB}{170,170,170}

\definecolor{brown}{HTML}{a52a2a}
\definecolor{darkcyan}{HTML}{0a888a}

\newcommand{\etal}{\textit{et~al.}\xspace}
\newcommand{\ie}{\textit{i.e.},\xspace}
\newcommand{\eg}{\textit{e.g.},\xspace}
\newcommand{\cf}{\textit{cf.}\xspace}

\newcommand{\wrt}{w.r.t.\xspace}

\usepackage{letltxmacro}
\newcommand{\code}[1]{\texttt{#1}}